\documentclass[aps,prd,superscriptaddress,amsfonts,amssymb,amsmath,showpacs,10pt]{revtex4-2}

\usepackage{silence}
\usepackage{bm}
\usepackage{amsfonts}
\usepackage{amssymb}
\usepackage{amsmath}
\usepackage{latexsym}
\usepackage[latin1]{inputenc}
\usepackage[T1]{fontenc}

\usepackage{graphicx}
\usepackage[caption=false]{subfig} % For subfigures with revtex4-2
\usepackage{booktabs}
\usepackage{dcolumn}
\usepackage{multirow}
\usepackage{hhline}

\usepackage[british]{babel}
\usepackage{textcomp}
\usepackage{ragged2e}
\usepackage{commath}
\usepackage[normalem]{ulem}

\usepackage{xcolor}
\usepackage{orcidlink}
\usepackage{epsfig}
\usepackage{xparse}
\renewcommand{\sout}[1]{}
\newcommand{\skm}[1]{{#1}}
\newcommand{\skmn}[1]{{#1}}

\newenvironment{skmcomment}
  {\par\begingroup}
  {\endgroup\par}
 
\allowdisplaybreaks[1]

\usepackage{hyperref}
\hypersetup{
    colorlinks=true,
    citecolor=blue,
    linkcolor=blue,
    filecolor=magenta,
    urlcolor=blue}
\begin{document}
%\color{red}
\color{black}       %% For one column

\title{Gauge invariant perturbations of \texorpdfstring{$F(T,T_G)$}{} Cosmology}

%\end{document}

\author{Shivam Kumar Mishra \orcidlink{0009-0006-4754-4103
}}
\email{shivamkumarmishra.mt@gmail.com}
\affiliation{Department of Mathematics,
Birla Institute of Technology and Science-Pilani, Hyderabad Campus, Jawahar Nagar, Kapra Mandal, Medchal District, Telangana 500078, India.}

\author{Jackson Levi Said\orcidlink{0000-0002-7835-4365}}
\email{jackson.said@um.edu.mt}
\affiliation{Institute of Space Sciences and Astronomy, University of Malta, Malta, MSD 2080}
\affiliation{Department of Physics, University of Malta, Malta}

 \author{B. Mishra\orcidlink{0000-0001-5527-3565}}
 \email{bivu@hyderabad.bits-pilani.ac.in}
 \affiliation{Department of Mathematics,
Birla Institute of Technology and Science-Pilani, Hyderabad Campus, Jawahar Nagar, Kapra Mandal, Medchal District, Telangana 500078, India.}

\begin{abstract}
The Gauss-Bonnet invariant connects foundational aspects of geometry with physical phenomena in a variety of ways. Teleparallel gravity offers a novel direction in which to use the Gauss-Bonnet invariant to go beyond standard cosmology. In this work, we explore the cosmological perturbations of teleparallel gravity generalized through the Gauss-Bonnet invariant. This is crucial in understanding the viability of these models beyond background analyses. We do this by taking a gauge-invariant approach, followed by popular gauge choices. It is essential to take this approach to understand the stability and healthiness of the underlying theory. We determine the equations of motion for all perturbative modes and provide a physical interpretation of the new contributions for each mode.
\end{abstract}

\maketitle

\section{Introduction} \label{sec:intro}

The last few decades have seen the $\Lambda$CDM model dominate as the best way to unify cosmological observations at different scales in a precise manner~\cite{Peebles:2002gy,Copeland:2006wr}. Here, the cosmological constant ($\Lambda$) dominates in the late-time accelerated expansion of the Universe~\cite{Riess:1998cb,Perlmutter:1998np}. At the same time, general relativity (GR) describes gravitational interactions and cold dark matter (CDM) primarily drives the primordial formation of large-scale structures while also being a stabilizing agent for galactic structures in the late Universe~\cite{Baudis:2016qwx,Bertone:2004pz}. The concordance, or $\Lambda$CDM, model features numerous fundamental questions about the nature of the Universe~\cite{Weinberg:1988cp}. However, putting aside these important open-questions, the $\Lambda$CDM model has increasingly been strained by the growing tension between observational constraints based on early time and late time cosmological surveys~\cite{CosmoVerseNetwork:2025alb,DiValentino:2020vhf,DiValentino:2020zio,DiValentino:2020vvd,Staicova:2021ajb}. At the same time, the problem of direct observational evidence for CDM continues to decrease in likelihood~\cite{LUX:2016ggv,Gaitskell:2004gd}, while the open questions in $\Lambda$CDM cosmology become more prominent in the community.

The literature is rich in proposed avenues for possible modifications to the $\Lambda$CDM model along with suggestions for different possible phenomena, where these models may offer their strongest deviations from the concordance model. This extensive family of cosmological models that expose different aspects of concordance cosmology include alternative behavior for CDM~\cite{Feng:2010gw,Dodelson:1993je,Joyce:2014kja,Abazajian:2012ys}, the addition of new dynamical features to dark energy~\cite{Copeland:2006wr,Benisty:2021gde,Benisty:2020otr,Bamba:2012cp}, extensions to the GR description of gravity~\cite{Clifton:2011jh,CANTATA:2021ktz,Bahamonde_2023,AlvesBatista:2021gzc,Addazi:2021xuf,Capozziello:2011et}, among many more. The approach these models take is primarily to add complexity to the $\Lambda$CDM in a given avenue. One interesting recourse is to reconsider part of the geometric foundations of the baseline $\Lambda$CDM model. Teleparallel gravity (TG) performs this task by interchanging the curvature associated with the Levi-Civita connection with the geometric torsion inferred from the teleparallel connection \cite{Bahamonde_2023,Krssak:2018ywd,Cai:2015emx}. This representation of gravitational interactions offers some advantages, including a well-defined energy-momentum tensor for the gravitational sector, a more dynamic relation to the equivalence principle \cite{Aldrovandi:2013wha}, as well as some positive indications from the observational sector \cite{Bahamonde_2023}.

The teleparallel connection produces TG as a curvature-free and metricity-satisfying formalism which hosts a limit that is dynamically equivalent to GR, called the teleparallel equivalent of general relativity (TEGR) \cite{Maluf:2013gaa,aldrovandi1995introduction}. The difference between TEGR and GR is borne out of a total divergence term $B$ in which the dynamical part is contained within the baseline torsion scalar $T$. This distinction plays a critical role in modifications of TEGR and possible differences in the infrared limit of the theory. A popular approach to probing physics beyond the $\Lambda$CDM model is to take a direct generalization of TEGR known as $f(T)$ gravity \cite{Ferraro:2006jd,Ferraro:2008ey,Bengochea:2008gz,Linder:2010py,Chen:2010va,Bahamonde:2019zea,Paliathanasis:2017htk,Farrugia:2020fcu,Bahamonde:2021srr,Bahamonde:2020bbc,Bahamonde:2022ohm,Bahamonde:2020lsm}, where the torsion scalar is considered arbitrarily in the gravitational action. Dissimilar to other approaches, $f(T)$ gravity produces generally second-order equations of motion, while also being consistent with a large spectrum of observational data. Another interesting extension to $\Lambda$CDM cosmology is to consider nontrivial contributions from the teleparallel equivalent of the Gauss-Bonnet scalar invariant $T_G$ \cite{Kofinas:2014daa,Kofinas:2014aka,Kofinas:2014owa}, which forms $F(T, T_G)$ gravity and has been shown to have advantageous properties in the cosmological context \cite{Bahamonde:2016kba,delaCruz-Dombriz:2017lvj,delaCruz-Dombriz:2018nvt}. This scalar invariant was first identified in Ref.~\cite{Kofinas:2014owa} where the field equations were presented. $F(T, T_G)$ was then applied to cosmology in Ref.~\cite{Kofinas:2014aka} where the Friedmann equations were derived and a formal dynamical systems analysis was considered for some extensions of TEGR. A reconstruction analysis was then produced in \cite{delaCruz-Dombriz:2017lvj,delaCruz-Dombriz:2018nvt}, where cosmologically advantageous models were identified. The class of models was further explored in Ref.~\cite{Bahamonde:2016kba}, where cosmological and astrophysical properties were studied more generally.

Additionally, there have been numerous important works on the study of $F(T, T_G)$ cosmology, focusing on its background evolution and its effects on physical processes in the primordial Universe. For instance, in Ref.~\cite{Pradhan:2023mjx}, a logarithmic extension of TEGR is considered using several compounded expansion data sets, yielding interesting results for the parameter space of this model. Along similar lines, Ref.~\cite{Balhara:2023mgj} undertakes an analogous constraint analysis for coupled mixed models, while Ref.~\cite{Gupta:2024qyn} studies the cosmological constraints for a Barrow-type dark energy fluid. Another important work on the constraints on model parameters in the $F(T, T_G)$ model space is Ref.~\cite{Pradhan:2025rir}, where cosmographic fits are also presented. There are numerous other works on the phenomenology of the underlying theory, including on the reconstruction of the functional form of the theory \cite{Dixit:2024qno}, the effects of the theory in the early-time processes of big bang nucleosynthesis \cite{Sultan:2025tws} and baryogenesis \cite{Majeed:2025tjp}, among others.

In this work, $F(T, T_G)$ is further studied through the prism of cosmological perturbations, which can expose possible ways in which the general class of models produce gravitational waves, couples to the large-scale structure of matter, as well as its healthiness in terms of internal stability and consistency with general aspects of cosmic evolution. This is critical in establishing the viability of possible $F(T, T_G)$ cosmological models. The tensor mode perturbations can be used both to consider multimessenger signals such as GW170817 \cite{TheLIGOScientific:2017qsa} and GRB170817A \cite{Goldstein:2017mmi} as well as potential B modes in the CMB \cite{Kamionkowski:2015yta}, while showing that vector perturbations diminish cosmologically is also essential for showing viability. Scalar perturbations expose the coupling of matter structures with gravity and have a particularly strong impact on CMB physics. Through the exploration of $F(T, T_G)$ cosmological perturbations, we aim to better understand the influence of these classes of models on the evolution of the Universe.\sout{We do this by considering the individual sector cosmological perturbations} \skmn{We do this by considering each sector of the cosmological perturbations} in a gauge-invariant way. This is then followed by four popular gauge illustrative case examples.

The work opens with a technical introduction in Sec.~\ref{sec:tech_intro} where the details of TG are discussed leading up to the introduction of $F(T, T_G)$ cosmology. This also contains the background equations of the class of models. In Sec.~\ref{sec:pert_back} cosmological perturbations are introduced in a gauge-invariant manner, which is then applied to $F(T, T_G)$ gravity in Sec.~\ref{sec:f_T_TG_pert}. The equations of motion are determined for each perturbative sector with an emphasis on the physical impacts of the modifications beyond the $\Lambda$CDM model. Example cases are illustrated in Sec.~\ref {sec:gauge_ch} for popular gauge choices. Finally, the work is summarized in Sec.~\ref{sec:conc}.

\section{\texorpdfstring{$F(T,T_G)$}{} Cosmology} \label{sec:tech_intro}

We begin by revisiting teleparallel gravity, outlining its physical foundations and examining its behavior within a maximally symmetric cosmological background. This overview establishes the groundwork for understanding its influence on cosmic evolution over time. Additionally, we emphasize the derivation of key equations from various assumptions and generalizations that govern the dynamics in this setting. This foundation paves the way for investigating perturbations around distinct backgrounds and studying structure formation within the teleparallel gravity formalism.

\subsection{\texorpdfstring{Teleparallel Gravity}{}}\label{sec:TG_int}

Gravity, due to its universality, has to be described by a field related to space-time itself. In Einstein's general theory of relativity, gravity is characterized by the most fundamental field, the metric, which defines the notion of distances and angles in the space-time manifold with a compatible Levi-Civita connection whose torsion vanishes, and curvature explains the changes due to gravity. In an equivalent description, called TEGR, the underlying structure is the same four-dimensional space-time manifold $\mathcal{M}$ \cite{Aldrovandi:2013wha}. \sout{For each point $p\in \mathcal{M}$ in a local chart $\{x^1,x^2,x^3,x^4\}$, the tangent space $T_p\mathcal{M}$ has basis  $\{\partial_1,\partial_2,\partial_3,\partial_4\}$. The dual space $T^*_p\mathcal{M}$ is spanned by $\{dx^1,dx^2,dx^3,dx^4\}$.  Greek indices such as $(\mu, \nu, \rho, \ldots)$ are used to denote coordinates and indices that belong to space-time. On the other hand, Latin indices $(a, b, c, \ldots)$ are assigned to coordinates in the tangent space. The tangent space is assumed to be locally Minkowskian, with the metric given by $\eta_{ab} = \mathrm{diag}(-1, 1, 1, 1)$. In this notation, the coordinates in space-time are indicated by $x^\mu$, while the coordinates in the tangent space are labeled $x^a$. These two sets of coordinates are considered to be related by invertible functions.} \skmn{For each point $p \in \mathcal{M}$ in a local chart $\{x^{\mu}\}$, the tangent space $T_{p}\mathcal{M}$ is a vector space attached to the manifold. We use Greek indices $(\mu, \nu, \rho, \dots)$ to denote holonomic indices related to the spacetime manifold, while Latin indices $(a, b, c, \dots)$ are assigned to a non-holonomic basis $\{e_{a}\}$ of the tangent space $T_{p}\mathcal{M}$. This basis is related to the coordinate basis through the components of the tetrad, $e_{a} = {e_{a}}^{\mu} \partial_{\mu}$.} When specialized to four dimensions, both Latin and Greek indices take values $ 0, 1, 2, 3$, with $x^0$ denoting the time coordinate.

Let $\{e_a(x)\}$ be a\sout{n arbitrary basis} \skmn{non-holonomic basis} for the tangent space $T_pM$ at the point $x$. The covariant derivative of the basis vector $e_a$ \sout{can be expressed as} is defined by \cite{Izumi:2012qj}
\begin{equation}
    \nabla e_a(x) = \omega^b{}_a(x)\, e_b(x)\,,\label{eq_1}
\end{equation}
where $\omega^b{}_a(x)$ is the connection one-form, called spin connection, \skmn{whose components are given by}
\begin{equation}
    \omega^b{}_a(x) = \langle e^b,\, \nabla e_a \rangle = \omega^b{}_{a\mu}(x)\, dx^\mu,\label{eq_2}
\end{equation}
Here, $\{e^b(x)\}$ denotes the coframe dual to $\{e_b(x)\}$, and $\{dx^\mu\}$ is the dual basis to the coordinate basis $\{\partial_\mu\}$. Similar to frame $e_a$, the coframe is related to the dual coordinate basis as $e^a = e^a\,_\mu\,dx^\mu$.

The spin connection is purely inertial, which means it keeps the special-relativistic role of representing inertial effects only. As a purely inertial connection, its curvature vanishes \cite{Aldrovandi:2013wha},

\begin{equation}
R^{a}\,_{b\mu\nu} = 2\partial _{[\mu }\omega ^{a}\,_ {|b|\nu ]} +2 \omega ^{a}\,_{c [\mu } \, \omega^{c}\,_{|b|\nu ]} = 0\, .\label{eq_3}
\end{equation}

The \sout{space-time metric is related to the tangent space Minkowski metric} \skmn{holonomic metric is related to the non-holonomic Minkowski metric} as
\begin{equation}
    g_{\mu\nu} = \eta_{ab}\, e^a\,_\mu\, e^b\,_\nu\,,\label{eq_4}
\end{equation}
which imposes orthonormality on the tetrads (The terms tetrads and vierbeins are used in the literature interchangeably, both meaning four legs, unlike vielbeins, which means many legs). Taking the determinant on both sides, one can see 
\begin{equation}
    e:=\det{(e^{a}\,_{\alpha})}=\sqrt{|g|}\,.\label{eq_5}
\end{equation}
Using the tetrad $e_a$, a \sout{general connection} \skmn{holonomic connection} $\Gamma^\rho\,_{ \nu\mu}$ can be related to the corresponding spin connection $\omega^a{}_{b \mu}$ \cite{Tseytlin:1982firstorder,Harst:2012tetrad,Aldrovandi:2013wha,Sadovski:2025holonomic}
\begin{equation}
    \Gamma^\rho\,_{ \nu\mu} = e_a\,^\rho \, \partial_\mu e^a\,_\nu + e_a\,^\rho \, \omega^a\,_{b \mu} e^b\,_\nu\,.\label{eq_6}
\end{equation}
The inverse relation is consequently given as,
\begin{equation}
    \omega^a\,_{b \mu} = e^a\,_\rho \, \partial_\mu e_b\,^\rho + e^a\,_\rho \, \Gamma^\rho\,_{ \nu\mu} e_b\,^\nu\,.\label{eq_7}
\end{equation}
Equations (\ref{eq_6}) and (\ref{eq_7}) illustrate that the \sout{total} \skmn{covariant} derivative of the tetrad acting on both indices is identically zero,
\begin{equation}
    \partial_\mu e^a\,_\nu - \Gamma^\rho\,_{ \nu\mu} e^a\,_\rho + \omega^a\,_{b \mu} e^b\,_\nu = 0\,.\label{eq_8}
\end{equation}
This implies that the tetrads are teleparallel, meaning they are covariantly constant $\skmn{\nabla_{X}^{\omega,\Gamma}}e_a = 0,$
where $\skmn{\nabla_{X}^{\omega,\Gamma}}$ denotes the covariant derivative along the vector $X$. A tedious but direct calculation \skmn{in components or just the realization that $R(\Gamma)=e^{*}R(\omega)$} shows that the teleparallel connection is flat 
\begin{equation}
R^{\lambda}\,_{\delta\mu\nu} = 2\partial _{[\mu }\Gamma ^{\lambda}\,_ {|\delta|\nu ]} +2 \Gamma ^{\lambda}\,_{\zeta [\mu } \, \Gamma ^{\zeta}\,_{|\delta|\nu ]} = 0\,. \label{eq_9}
\end{equation}

It is important to note that for a given metric \sout{defined by Eq. \eqref{eq_4}}, the associated vierbein is not uniquely determined \skmn{as Eq. \eqref{eq_4} is a non-invertible field transformation and vierbeins leading to the same metric are Lorentz transformations of each other.} To maintain covariance of the theory, one must appropriately select the spin connection to compensate for the chosen vierbein. Furthermore, a straightforward application of Minkowskian geometry may introduce spurious inertial effects. Specifically, as demonstrated in Eq. \eqref{eq_4}, when the global space-time is Minkowski, this condition is satisfied only if the vierbeins exactly correspond to local Lorentz transformations \cite{Aldrovandi:2013wha}. \sout{To maintain consistency within flat space-times, it is helpful to define} \skmn{Owing to the flatness condition \eqref{eq_3}, the inertial spin connection can be written as}

\begin{equation}
    \omega^{a}\,_{b\mu} := \varLambda^{a}\,_{c} \, \partial_{\mu} \varLambda_{b}\,^{c}\,.\label{eq_10}
\end{equation}

Here, the symbol $\varLambda^{a}\,_{c}$ represents Lorentz transformations, including boosts and rotations. One can thus \sout{always} choose a Lorentz frame in which the spin connection vanishes. Tetrads that satisfy this property are called proper tetrads, and the corresponding set-up is referred to as the Weitzenb\"ock gauge.

Since the basis vectors $\{e_a\}$ are not integrable in general, the teleparallel connection \eqref{eq_6} has torsion whose components are given by 
\begin{equation}
  T^{\lambda}\,_{\mu\nu} 
=\, \Gamma^{\lambda}\,_{\nu\mu} - \Gamma^{\lambda}\,_{\mu\nu}\,.\label{eq_11}  
\end{equation}
The tetrad field can also be utilized to define a \sout{torsion-free affine connection} \skmn{torsion-free metric compatible affine connection}, denoted by $\bar{\Gamma}^\rho\,_{\mu\nu}$, which is the Levi-Civita connection uniquely determined by the metric given in Eq. \eqref{eq_4}. Its components are given as

\begin{equation}
    \bar{\Gamma}^\lambda\,_{\mu\nu} = \frac{1}{2} g^{\lambda\rho} \left( \partial_\mu g_{\rho\nu} + \partial_\nu g_{\rho\mu} - \partial_\rho g_{\mu\nu} \right)\,.\label{eq_12}
\end{equation}
The Levi-Civita and teleparallel connections are related by a tensor called the Contortion tensor
\begin{equation}
  {\Gamma}^\lambda\,_{\mu\nu} =  \bar{\Gamma}^\lambda\,_{\mu\nu}+\mathcal{K}^\lambda\,_{\mu\nu}\,.\label{eq_13} 
\end{equation}
Using the metricity of Levi-Civita connection and Eqs. \eqref{eq_11} and \eqref{eq_13} it is straightforward to deduce
\begin{equation}
 \mathcal{K}^{\lambda }\,_{\mu\nu}= \frac{1}{2}(T_{\mu}\,^{ \lambda}\,_{\nu}+T_{\nu}\,^{\lambda }\,_\mu-T^{\lambda }\,_{\mu\nu})\,.\label{eq_14}   
\end{equation}
Using the flatness of the teleparallel connection and the relation \eqref{eq_13}, one can write the Riemann tensor for the Levi-Civita connection as
\begin{equation}
\bar{R}{}^{\mu}\,_{\nu\lambda\sigma} =
\mathcal{K}^{\mu}\,_{\tau\sigma} \mathcal{K}^{\tau}\,_{\nu\lambda}
- \mathcal{K}^{\mu}\,_{\tau\lambda} \mathcal{K}^{\tau}\,_{\nu\sigma}
+ \bar{\nabla}_{\sigma} \mathcal{K}^{\mu}\,_{\nu\lambda}
- \bar{\nabla}_{\lambda} \mathcal{K}^{\mu}\,_{\nu\sigma}\,.
\label{eq_15}
\end{equation}
From this relation, the Ricci scalar can be expressed as 
\begin{equation}
\bar{R} = \mathcal{K}^{\mu}\,_{\lambda \nu} \mathcal{K}^{\lambda \nu}\,_{ \mu}
- \mathcal{K}^{\mu}\,_{\lambda \mu} \mathcal{K}^{\lambda \nu}\,_{ \nu}
+ 2 \bar{\nabla}_{\mu} \mathcal{K}^{\mu \nu}\,_{ \nu}
= -T + 2 \bar{\nabla}_{\mu} T^{\nu \mu}\,_{ \nu}\,,
\label{eq_16}
\end{equation}
where the torsion scalar $T$ is defined as 
\begin{equation}
\begin{aligned}
T &:=\mathcal{K}^{\mu}\,_{\lambda \mu} \mathcal{K}^{\lambda \nu}\,_{ \nu} -  \mathcal{K}^{\mu}\,_{\lambda \nu} \mathcal{K}^{\lambda \nu}\,_{\mu} \\
&= \frac{1}{4} T^{\lambda \mu \nu} T_{\lambda \mu \nu} + \frac{1}{2} T^{\lambda \mu \nu} T_{\nu \mu \lambda} - T_{\mu}\,^{\mu \lambda} T^{\nu}\,_{\nu \lambda}\,.\label{eq_17}
\end{aligned}
\end{equation}
An intriguing aspect of the torsion scalar  $T$  is that it can be expressed in a form involving the so-called superpotential $ S_{\mu}\,^{\nu \lambda}$ as \cite{Bahamonde_2023}
\begin{equation} 
T =  S_{\mu}\,^{\nu \lambda} T^{\mu}\,_{\nu \lambda}\,.\label{eq_18}
\end{equation}
The superpotential tensor is defined as
\begin{equation} 
S_{\mu}\,^{\nu \lambda} :=\frac{1}{2}( \mathcal{K}^{\nu \lambda}\,_{\mu} - \delta_{\mu}\,^{\nu} T_{\sigma}\,^{\sigma \lambda} + \delta_{\mu}\,^{\lambda} T_{\sigma}\,^{\sigma \nu})\,.\label{eq_19}
\end{equation}
% which has the antisymmetry property
% \begin{equation} \label{eq:antisymmetry}
% S_{\mu}{}^{\nu \lambda} = - S_{\mu}{}^{\lambda \nu}.
% \end{equation}
 From Eq. \eqref{eq_16}, it is easy to see that 
 \begin{equation}
e\, \bar{R} = -e\,T + 2\partial_{\mu}\left(e\, T_{\nu}\,^{ \mu\nu}\right)\,.\label{eq_20}
\end{equation}
Therefore, the action constructed from the torsion scalar $T$ corresponds to TEGR, which is dynamically equivalent to general relativity (GR). Inspired by curvature-based gravity theories, TEGR can be generalized to an $f(T)$ gravity framework where the Lagrangian is replaced by a general function $f(T)$. A significant advantage of $f(T)$ gravity compared to its curvature-based counterpart $f(\bar{R})$ is that the resulting field equations are second order in derivatives of the tetrad fields, making the equations of motion more tractable.

Following a similar approach, i.e., rewriting quantities associated with the Levi-Civita connection in terms of those related to the teleparallel connection up to boundary terms, Ref. \cite{Kofinas:2014owa} introduced the teleparallel counterpart of the Gauss-Bonnet invariant 
$\bar{G}=\bar{R}^{2}-4\bar{R}_{\mu\nu}\bar{R}^{\mu\nu}+\bar{R}_{\mu\nu\lambda\rho}\bar{R}^{\mu\nu\lambda\rho}$, 
which is described by a novel torsion-based scalar $T_{G}$. They also derived the equations of motion for the modified gravity theory defined by the function $F(T, T_{G})$. The relation between these quantities is given by 
\begin{equation}
e\bar{G} = eT_{G} + \text{total divergence}\,,
\label{eq_21}
\end{equation}
where $\bar{G}$ denotes the Gauss-Bonnet term constructed from the Levi-Civita connection, and 
\begin{equation}
   T_G = \big(\mathcal{K}^{j}\,_{ea}\mathcal{K}^{ek}\,_{b}
\mathcal{K}^{l}\,_{fc}\mathcal{K}^{fm}\,_{d} 
- 2\mathcal{K}^{jk}\,_{a}\mathcal{K}^{l}\,_{eb}
\mathcal{K}^{e}\,_{fc}\mathcal{K}^{fm}\,_{d} 
 + 2\mathcal{K}^{jk}\,_{a}\mathcal{K}^{l}\,_{eb}
\mathcal{K}^{em}\,_{f}\mathcal{K}^{f}\,_{cd} 
+ 2\mathcal{K}^{jk}\,_{a}\mathcal{K}^{l}\,_{eb}\,\,
\partial_d\mathcal{K}^{em}\,_{c}\big)\delta^{abcd}_{jklm}\,,
\label{eq_22} 
\end{equation}
where the generalized delta $\displaystyle\delta^{abcd}_{jklm}$ is the determinant tensor built from Kronecker deltas. It should be noted that the indices labeled by Latin letters correspond to Lorentz indices. Hence, $T_G$ serves as the teleparallel equivalent of $\bar{G}$ in the sense that, for arbitrary dimension $n$, the action
\begin{equation}
S_{T_G} =  \int_{\mathcal{M}} d^{n}x\, e\, T_G\,,
\label{eq_23}
\end{equation}
when varied with respect to the $n$-dimensional tetrads (vielbeins), it yields precisely the same field equations as the action
\begin{equation}
S_{\bar{G}} = \int_{\mathcal{M}} d^{n}x\, e\, \bar{G}\,,
\label{eq_24}
\end{equation}
varied with respect to the metric in $n$-dimensions.

Since the scalar $T_G$ is also a topological invariant in four dimensions, similar to the Gauss-Bonnet term $\bar{G}$, actions that depend linearly on $T_G$ do not lead to new dynamics and are therefore not of primary interest. As a result, a broader class of gravitational theories is considered where the action is given by a general functional dependence on both the torsion scalar $T$ and the teleparallel Gauss-Bonnet scalar $T_G$, namely
\begin{equation}
S_{TGB} =  \int_{\mathcal{M}}  d^{n} x\, e\, F(T, T_G)\,,
\label{eq_25}
\end{equation}
where $F$ is an arbitrary function of $T$ and $T_G$. This framework represents a novel extension of TG, distinct from both the simpler $F(T)$ theories and the well-known curvature-based $F(\bar{R},\bar{G})$ gravity theories. It generalizes TEGR, which is recovered by choosing $F(T, T_G) = -T$, while the standard Einstein-Gauss-Bonnet gravity corresponds to the particular choice $F(T, T_G) = -T + \alpha T_G$, where $\alpha$ is the Gauss-Bonnet coupling constant. This construction thus allows for richer gravitational dynamics by including both torsion and higher-order torsion contributions beyond linear terms.

\skmn{Since tetrads are fundamental variables in the theory, f}ield equations from the variation of action \eqref{eq_25} with respect to $n$-dimensional tetrads in differential form language was given in original work \cite{Kofinas:2014owa}, both in covariant form as well as in Weitzenb\"ock gauge. We present the field equations in a way that makes the reduction to the $F(T)$ case simple and the dynamical aspects of $T_G$ clearer. In $n=4$, field equations of $F(T, T_G)$ gravity are given as \cite{Bahamonde:2016kba,Bahamonde_2023} 
\begin{alignat}{2}
& \: & &\mathbb{E}_a\,^{\mu}:=\frac{1}{e}F_{T}\partial_{\nu}(e S_{a}\,^{\mu\nu}) + S_{a}\,^{\mu\nu} (\partial_{\nu}F_{T}) + F_{T}T^{b}\,_{\nu a}S_{b}\,^{\mu\nu} +\frac{1}{4}F e_a\,^\mu -  \frac{1}{2}F_{T_G} \delta^{mbcd}_{ijk\ell} e_{d}\,^{\mu} \mathcal{K}^{ij} {} _m \mathcal{K}^{k} {} _ {eb} \partial_a \mathcal{K}^{e\ell} {} _c  \nonumber\\[0.5ex]
& \: & &+\frac{1}{4e}\partial_{\beta}\Big(\eta_{a\ell}(Y^{b[\ell h]} - Y^{h[\ell b]} + Y^{\ell[b h]})e_{h}\,^{\beta} e_{b}\,^{\mu}\Big) + \frac{1}{4e} T_{iab} e_{h}\,^{\mu} ( Y^{b[i h]} - Y^{h[i b]} + Y^{i[b h]} )  = 0\,,
\label{eq_26}
\end{alignat}
\skmn{where $F_T$ and $F_{T_G}$ are partial derivatives of $F$ with respect to $T$ and $T_G$ respectively and} the tensor $ Y^{b}{}_{ij} $ is defined as
\begin{align}
    Y^{b}{}_{ij} &:= e F_{T_{G}} X^{b}{}_{ij} - 2 \delta^{cabd}_{elkj} \partial_{\mu} \big( e F_{T_G} e_{d}\,^{\mu} \mathcal{K}^{el}\,_{c} \mathcal{K}^{k}\,_{ia} \big) \,,
\label{eq_27}
\end{align}
with
\begin{alignat}{2}
X^{a}{}_{ij} &:= \,\frac{\partial T_G}{\partial \mathcal{K}_a {}^ {ij}}\,=\mathcal{K}_j {}^e {} _b \mathcal{K}^k {} _ {fc} \mathcal{K}^{fl} {} _d \delta^{abcd} _ {iekl} +  
 \mathcal{K}^e {} _ {ib} \mathcal{K}^k {} _ {fc} \mathcal{K}^{fl} {} _d \delta^{bacd} _ {ejkl} + 
 \mathcal{K}^k {} _ {ec} \mathcal{K}^{ef} {} _b \mathcal{K}_j {}^l {} _d \delta^{cbad} _ {kfil} + 
 \mathcal{K}^f {} _ {ed} \mathcal{K}^{el} {} _b \mathcal{K}^k {} _ {ic} \delta^{dbca} _ {flkj} \nonumber\\[0.5ex]
& \:\hspace{5em} + 2 \mathcal{K}^k {} _ {eb} \mathcal{K}^{el} {} _f \mathcal{K}^f {} _ {cd} \delta^{abcd} _ {ijkl}-2 \mathcal{K}^{k}{}_{e b} \mathcal{K}^{e}{}_{f c} \mathcal{K}^{f l}{}_{d} \delta_{i j k l}^{a b c d}
- 2 \mathcal{K}^{k e}{}_{b} \mathcal{K}^{j}{}_{f c} \mathcal{K}^{f l}{}_{d} \delta_{k e i l}^{b a c d}
- 2 \mathcal{K}^{e f}{}_{c} \mathcal{K}^{k}{}_{i b} \mathcal{K}^{j l}{}_{d} \delta_{e f k l}^{c b a d}
 \nonumber\\[0.5ex]
& \:\hspace{5em} - 2 \mathcal{K}^{f l}{}_{d} \mathcal{K}^{k}{}_{e b} \mathcal{K}^{e}{}_{i c} \delta_{f l k j}^{d b c a}
 + 
 2 \mathcal{K}^{ke} {} _b \mathcal{K}_j {}^l {} _f \mathcal{K}^f {} _ {cd} \delta^{bacd} _ {keil} + 
 2 \mathcal{K}^{el} {} _f \mathcal{K}^k {} _ {ib} \mathcal{K}^a {} _ {cd} \delta^{fbcd} _ {elkj} + 
 2 \mathcal{K}^{fc} {} _d \mathcal{K}^{d} {} _ {eb} \mathcal{K}^{el} {} _i \eta_{mj} \delta^{dbma} _ {fckl}  \nonumber\\[0.5ex]
& \: \hspace{5em} + 2 \mathcal{K}^k {} _ {eb} \delta^{abcd} _ {ijkl} \partial_d \mathcal{K}^{el} {} _c + 
 2 \mathcal{K}^{ke} {} _b \delta^{bacd} _ {keil} \partial_d \mathcal{K}_j{}^l {} _c \,.
\label{eq_28}
\end{alignat}
Before studying cosmology, we note that using the Weitzenb\"ock gauge in $F(T, T_G)$ gravity breaks local Lorentz invariance by selecting specific frames. While TEGR remains Lorentz covariant, the equations in $f(T)$ and $F(T, T_G)$ formulations are only \sout{form} \skmn{diffeomorphism}-invariant. This does not invalidate the theory, as solutions still uniquely determine the metric, though not all tetrads satisfy the field equations. Hence, it is helpful to decompose the equations into their symmetric and antisymmetric components, since the antisymmetric part accounts for frame-dependent effects and matches the equations obtained from varying the flat spin connection in covariant formulations \cite{Golovnev:2017}. 

For the sake of brevity, we do not write the symmetric and antisymmetric parts of equation (\ref{eq_26}) explicitly but indicate a general procedure to do it in generalized TG models. We have $\displaystyle\frac{\delta S_{TGB}}{e\,\delta e^a\,_{\mu}}=\mathbb{E}_a\,^{\mu}$, use metric and tetrads to get equations in \sout{global indices} \skmn{holonomic indices} as $ \mathbb{E}_{\nu\mu}=g_{\mu\rho}\,e^a\,_{\nu}\,\mathbb{E}_a\,^{\rho}$ and split the symmetric and antisymmetric part as 
\begin{eqnarray}
    \mathbb{E}_{\nu\mu} = \mathbb{E}_{(\nu\mu)} + \mathbb{E}_{[\nu\mu]} \label{eq_29}
\end{eqnarray}
We also comment on the general structure of the field equations in generalized TG models. The breaking of local Lorentz symmetry means the \sout{Bianchi identities} \skmn{energy momentum conservation} no longer holds automatically \cite{NGR}. Defining the energy-momentum tensor for any matter action $S_m$ by
\begin{equation}
\frac{\delta S_m}{\delta e^a\,_{\mu}} =  e \,e_a\,^\nu\,\Theta_\nu\,^{\mu}.\label{eq_30}
\end{equation}
Thus, for minimally coupled theories, one obtains $\mathbb{E}_a\,^{\mu}\propto e_a\,^\nu\,\Theta_\nu\,^{\mu}$. The invariance of the matter action under diffeomorphisms, $e^a\,_{\mu} \to e^a\,_{\mu} + \mathcal{L_\zeta}\,e^a\,_{\mu}$ implies
\begin{equation}
\frac{1}{ e } \partial_{\mu} \big(  e\,  {\Theta}_{\nu}\,^{\mu} \big) - {\Theta}_{\alpha}\,^{\beta} e_{a}\,^{\alpha} \partial_{\nu} e^{a}\,_{\beta} = 0\,,\label{eq_31}
\end{equation}
which can be expressed as
\begin{equation}
\bar{\nabla}_{\mu} {\Theta}_\nu\,^{\mu} - \mathcal{K}^{\alpha\beta}\,_\nu {\Theta}_{\alpha\beta} = 0\,.\label{eq_32}
\end{equation}

When local Lorentz invariance holds, invariance under local frame rotations $e^{a}\,_{\mu} \to \varLambda^{a}\,_{ b} e^{b}\,_{\mu}$ forces ${\Theta}^{\mu\nu}$ to be symmetric. Because the contortion tensor is antisymmetric in the first two indices, the usual \sout{Bianchi identities} conservation follows. This symmetry is lost in generalized teleparallel gravity models, but the vanishing of the antisymmetric part of the field equations requires the antisymmetric component of ${\Theta}^{\mu\nu}$ to vanish, thereby restoring the \sout{Bianchi identities} \skmn{Lorentz invariance}. Therefore the condition
\begin{equation}
     \mathbb{E}_{[\mu\nu]}=0 \quad\forall\;\; \mu,\nu\label{eq_33}
\end{equation}
is imposed. Interestingly, the above condition for the maximally symmetric ansatz is trivially satisfied. For perturbations, it is used to constrain Lorentz variables. We now proceed to the cosmological background.
\subsection{Background Cosmology}
    In this part, we explore the application of $F(T, T_G)$ gravity within a cosmological background. To begin, we incorporate the matter sector alongside the gravitational part by defining the total action in $4$-dimensions as
\begin{eqnarray}
S_{total} = \frac{1}{2\kappa^{2}} \int_{\mathcal{M}} d^{4}x\, e\, F(T,T_G) + S_m\,,
\label{eq_34}
\end{eqnarray}
where $S_m$ corresponds to the matter EM-tensor $\Theta_{\mu\nu}$, which will appear on RHS of Eq. \eqref{eq_26} as $e_a\,^\nu\,\Theta_\nu\,^{\mu}$and $\kappa^{2} = 8\pi G$ denotes the four-dimensional Newton coupling. Next, aiming to study the cosmological consequences of this action, we adopt a spatially flat cosmological metric ansatz
\begin{equation}
ds^{2} = - dt^{2} + a^{2}(t) \delta_{\hat{i}\hat{j}} dx^{\hat{i}} dx^{\hat{j}}\,,
\label{eq_35}
\end{equation}
where $a(t)$ is the scale factor and the indices with hats run over the three spatial dimensions. This metric can be derived from the diagonal vierbein
\begin{equation}
\left[e^{a}\,_{\mu}\right] = \mathrm{diag}(1, a(t), a(t), a(t))\,,
\label{eq_36}
\end{equation}
via relation \eqref{eq_4}. Its dual vierbein is given by
$\;\left[e_{a}\,^{\mu}\right] = \mathrm{diag}(1, a^{-1}(t), a^{-1}(t), a^{-1}(t))$ and the determinant evaluates to $e = a^{3}(t)$. Substituting the vierbein \eqref{eq_36} into definitions \eqref{eq_17} and \eqref{eq_22} we obtain
\begin{eqnarray}
\label{eq_37}
T &=& 6 \frac{\dot{a}^{2}}{a^{2}} = 6 H^{2}\,, \\
\label{eq_38}
T_G &=& 24 \frac{\dot{a}^{2}}{a^{2}} \frac{\ddot{a}}{a} = 24 H^{2} \left(\dot{H} + H^{2}\right)\,,
\end{eqnarray}
where $H = \displaystyle\frac{\dot{a}}{a}$ is the Hubble parameter and dot denotes derivative with respect to time $t$. Furthermore, inserting \eqref{eq_36} into the general field equations and performing straightforward manipulations lead to the modified Friedmann equations \cite{Kofinas:2014daa}
\begin{equation}
F - 12 H^{2} F_{T} - T_{G} F_{T_{G}} + 24 H^{3} \dot{F_{T_{G}}} = 2 \kappa^{2} \rho\,,
\label{eq_39}
\end{equation}
\begin{eqnarray}
&& F -4H\dot{F_{T}}- 4(\dot{H} + 3 H^{2}) F_{T}  - T_{G} F_{T_{G}} + \frac{2}{3H} T_{G} \dot{F_{T_{G}}}+8H^{2}\ddot{F_{T_{G}}}  = -2 \kappa^{2} p\,,
\label{eq_40}
\end{eqnarray}
 with the right-hand side originating from the variation of the matter action, described by a perfect fluid with energy density $\rho$ and pressure $p$. i.e. 
\begin{equation}
    \Theta_{\mu\nu}=(p+\rho)\mathrm{n}_\mu\,\mathrm{n}_\nu+p\,g_{\mu\nu}\label{eq_41}
\end{equation}
subject to the condition, $\mathrm{n}^\mu\,\mathrm{n}_\mu=-1$.
Due to the homogeneity and isotropy of the background, quantities can only depend on time. Therefore, we have $\displaystyle\Theta_{00} = \rho$, $\displaystyle\Theta_{\hat{i}\hat{j}} = {p}\,a^{2} \delta_{\hat{i}\hat{j}}$, and $\Theta^{\hat{i}}_{\,\,\hat{i}} = 3p$. Indices with a hat on them are spatial, i.e., $\hat{i}=1,2,3.$
In these equations, time derivatives of $F_T$ and $F_{T_G}$ are expanded as $\displaystyle \dot{F_{T}} = F_{TT} \dot{T} + F_{T T_G} \dot{T}_G\,$, 
$\displaystyle\dot{F_{T_{G}}} = F_{T T_G} \dot{T} + F_{T_G T_G} \dot{T}_G\,$, and 
$\displaystyle\ddot{F_{T_{G}}} = F_{T T T_G} \dot{T}^{2} + 2 F_{T T_G T_G} \dot{T} \dot{T}_G + F_{T_G T_G T_G} \dot{T}_G^{2} + F_{T T_G} \ddot{T} + F_{T_G T_G} \ddot{T}_G\,$, where subscripts such as $F_{TT}$, $F_{T T_G}$ denote \sout{partial derivatives of $F_T$ with respect to $T$ and $T_G$} \skmn{2nd order partial derivatives of $F$}. The quantities $\dot{T}$, $\ddot{T}$, $\dot{T}_G$ and $\ddot{T}_G$ are obtained by differentiating Eqs. \eqref{eq_37} and (\ref{eq_38}) with respect to time.

 The background matter components are conserved independently, satisfying the continuity equation
\begin{eqnarray}
 \bar{\nabla}_\mu\,\Theta_\nu\,^{\mu}\,&=&\,0\,, \label{eq_42} \\[1.5ex]
 \,\,{\implies} \dot{\rho}+3H(\rho+p)&=&0\,.\qquad(\nu=0)\, \label{eq_43}  
\end{eqnarray}
For cases $\nu=1,2,3$, LHS of (\ref{eq_42}) vanishes identically at background. This is not the case with the perturbed EM-tensor. The first order expansions of components of perturbed conservation equations are given in Appendix  \ref{app_B}.

\section{Gauge Invariant Cosmological Perturbations} \label{sec:pert_back}

Cosmological perturbations provide insights into aspects of the Universe not evident in the background cosmology, including the formation of cosmic structures and the origin of the gravitational wave background. Cosmological perturbation theory involves decomposing fundamental field perturbations into components that are irreducible \skmn{under the stabilizer group of the background, which, for FLRW, is the group of spatial rotations}. This is achieved by separating temporal and spatial parts, splitting into trace and trace-free components, and further dividing spatial parts into divergent and divergence-free elements. 

\begin{skmcomment}
      A spatial covariant vector $\mathrm{V}_{\hat{i}}$ can be decomposed into scalar and divergenceless vector parts as
\begin{align}\label{eq_44}  
\mathrm{V}_{\hat{i}}=\partial_{\hat{i}}\mathrm{V}+\mathrm{V}_{\hat{i}}\,
\end{align}
with the transverse component satisfying
\begin{align}\label{eq_45}  
\partial^{\hat{i}}\mathrm{V}_{\hat{i}}=0\,.
\end{align}
Since any spatial covariant antisymmetric tensor field $\mathrm{A}_{\hat{i}\hat{j}}$ can always be Hodge-dual to a covariant vector through
\begin{align}\label{eq_46}  
\mathrm{V}_{\hat{i}}=\frac{1}{2}\epsilon_{\hat{i}\hat{j}\hat{k}}\mathrm{A}^{\hat{j}\hat{k}}\,,
\end{align}
the antisymmetric tensor $\mathrm{A}_{\hat{i}\hat{j}}$ can be written as
\begin{align}\label{eq_47}  
\mathrm{A}_{\hat{i}\hat{j}}=\epsilon_{\hat{i}\hat{j}\hat{k}}\left(\partial^{\hat{k}}\mathrm{V}+\mathrm{V}^{\hat{k}}\right)\,.
\end{align}
A symmetric spatial tensor $\mathrm{S}_{\hat{i}\hat{j}}$ admits the scalar--vector--tensor decomposition
\begin{align}\label{eq_48}  
\mathrm{S}_{\hat{i}\hat{j}}=\hat{\mathrm{S}}\delta_{\hat{i}\hat{j}}
+\partial_{\hat{i}}\partial_{\hat{j}}\mathrm{S}
+2\partial_{(\hat{i}}\mathrm{S}_{\hat{j})}
+\hat{\mathrm{S}}_{\hat{i}\hat{j}}\,
\end{align}
where parentheses denote symmetrization, and the last term represents the traceless and transverse tensor contribution. Finally, any second-rank spatial tensor $\mathrm{T}_{\hat{i}\hat{j}}$ can be expressed as the sum of symmetric and antisymmetric parts, yielding
\begin{align}
\mathrm{T}_{\hat{i}\hat{j}}&=\mathrm{S}_{\hat{i}\hat{j}}+\mathrm{A}_{\hat{i}\hat{j}}\nonumber \\
&=\hat{\mathrm{S}}\delta_{\hat{i}\hat{j}}
+\partial_{\hat{i}}\partial_{\hat{j}}\mathrm{S}
+2\partial_{(\hat{i}}\mathrm{S}_{\hat{j})}
+\epsilon_{\hat{i}\hat{j}\hat{k}}\left(\partial^{\hat{k}}\mathrm{V}+\mathrm{V}^{\hat{k}}\right)
+\hat{\mathrm{S}}_{\hat{i}\hat{j}}~.\label{eq_49}  
\end{align}

\end{skmcomment}

Since the tetrad is the fundamental dynamical variable in our formulation, we perturb it as 
\begin{equation}
    e^a\,_\mu\,=\, \mathring{e}{^a\,_\mu}+\delta\, e^a\,_\mu\label{eq_50}
\end{equation}
where $\mathring{e}{^a\,_\mu}$ is the background tetrad i.e. diagonal FLRW tetrad
\begin{equation}
   \left[\mathring{e}{^a\,_\mu}\right] \,=\,\mathrm{diag}(1, a(t), a(t), a(t))\,.\label{eq_51}
\end{equation}
 \skm{Note that $e^{0}_{~0}$ behaves as a spatial scalar quantity, 
while $e^{0}_{~\hat{i}}$ and $\delta_{\hat{i}\hat{j}}\delta_{a}^{~\hat{j}}e^{a}_{~0}$ transform as spatial covariant vector fields, 
and $\delta_{\hat{j}\hat{k}}\delta_{a}^{~\hat{k}}e^{a}_{~\hat{i}}$ transforms as a spatial covariant rank-two tensor field, \skmn{under spatial rotations.} Hence }the perturbed tetrad \skmn{having 16 degrees of freedom (dof)} is given by \cite{Bahamonde_2023}
\begin{equation}
    \left[\delta\, e^{a}\,_{\mu}\right] = \left[\begin{array}{cc}
     \phi & a\left(\partial_{\hat{i}}\beta+\beta_{\hat{i}}\right) \\
    \delta^{\tilde{l}}{}_{\hat{i}}\left(\partial^{\hat{i}}B+B^{\hat{i}}\right) & a\delta^{\tilde{l}\,\hat{i}}\left(\psi\delta_{\hat{i}\hat{j}}+\partial_{\hat{i}}\partial_{\hat{j}}E + 2\partial_{(\hat{i}}h_{\hat{j})}+\frac{1}{2}h_{\hat{i}\hat{j}}+\epsilon_{\hat{i}\hat{j}\hat{k}}\left(\partial^{\hat{k}}\sigma+\sigma^{\hat{k}}\right)\right)
    \end{array}\right]\,, \label{eq_52}
\end{equation}
where perturbations are categorised as follows under 3-rotations:
\begin{enumerate}
    \item Scalars: $ \phi$, $\beta$, $B$, $\psi$, $E$ (5 dof)
    \item Pseudoscalar: $\sigma$ (1 dof)
    \item Vectors: $\beta_{\hat{i}}$, $B_{\hat{i}}$, $h_{\hat{i}}$ (solenoidal, 6 dof)
    \item Pseudovector: $\sigma_{\hat{i}}$ (solenoidal, 2 dof)
    \item Tensor: $h_{\hat{i}\hat{j}}$ (symmetric, transverse, traceless, 2 dof)
\end{enumerate}
Thus, a total of 16 dof are recovered, because \sout{unlike metric perturbations, tetrad perturbations} \skmn{unlike metric, tetrads} need not be symmetric. The additional six degrees of freedom arise from the Lorentz group, under which the metric remains invariant. The perturbed tetrad \eqref{eq_52} preserves the metric's symmetries and maintains the Weitzenb\"ock gauge at the perturbative level \cite{Bahamonde:2020lsm}. That is, its spin connection components remain consistent with the cases with vanishing values, ensuring the tetrad is appropriate for perturbative analysis. 

The corresponding perturbed metric takes the form
\begin{equation}
\left[\delta g_{\mu\nu}\right] = \left[\begin{array}{cc}
-2 \phi & a\left(\partial_{\hat{i}}(B-\beta)+(B_{\hat{i}}-\beta_{\hat{i}})\right)\\
a\left(\partial_{\hat{i}}(B-\beta)+(B_{\hat{i}}-\beta_{\hat{i}})\right) & 2a^{2}\left(\psi\delta_{\hat{i}\hat{j}}+\partial_{\hat{i}}\partial_{\hat{j}}E+2\partial_{(\hat{i}}h_{\hat{j})}+\frac{1}{2}h_{\hat{i}\hat{j}}\right)
\end{array}\right]\,, \label{eq_53}
\end{equation}
We emphasize that only vectors and pseudovectors can interact through terms such as \cite{Izumi:2012qj}
\begin{equation}
\epsilon_{\hat{i}\hat{j}\hat{k}} \left(\partial_{\hat{i}} \beta_{\hat{j}} \right) \sigma_{\hat{k}}\,.
\label{eq_54}
\end{equation}
When parity is conserved, no other mode couplings occur. Thus, except for vector and pseudovector modes, the different perturbation modes can be analyzed independently. At linear order, none of the modes are coupled with any other and thus are studied separately.

Now coming to perturbations of the energy-momentum tensor, imposing that the complete perturbed energy-momentum tensor represents a slight deviation from the background configuration guarantees that it continues to possess a unique timelike eigenvector  $u^\mu= \mathring{u}^\mu+\delta u^\mu$, still normalized as $u^\mu\,u_\mu=-1$, with $\mathring{u}^\mu=\mathrm{n}^\mu$. \skm{This allows us to adopt the standard decomposition for a perfect fluid accounting for fluctuations in energy density $\delta\rho$, pressure $\delta p$ and the fluid velocity vector $\delta u^\mu$, ensuring that the conservation laws $\bar{\nabla}_{\nu}\Theta_{\mu}\,^{\nu}=0$ are consistently satisfied at the linear order where $\Theta_{\mu\nu}=\mathring{\Theta}_{\mu\nu}+\delta \Theta_{\mu\nu}$}. Thus, the perturbed EM-tensor is given by:

\begin{equation}
\left[\delta \Theta_{\mu\nu}\right] = \left[\begin{array}{cc}
\delta \rho + 2 \rho  \phi & a \rho \left(\partial_{\hat{i}} (B - \beta) + (B_{\hat{i}} - \beta_{\hat{i}})\right) - a (\rho + p) \, \delta u_{\hat{i}} \\
a \rho \left(\partial_{\hat{i}} (B - \beta) + (B_{\hat{i}} - \beta_{\hat{i}})\right) - a (\rho + p) \, \delta u_{\hat{i}} & a^{2}\left(\delta p \,  + 2 p \left(\psi \delta_{\hat{i}\hat{j}} + \partial_{\hat{i}} \partial_{\hat{j}} E + 2 \partial_{(\hat{i}} h_{\hat{j})} + \frac{1}{2} h_{\hat{i} \hat{j}}\right)\right)
\end{array}\right]
\label{eq_55}
\end{equation}
$\delta u_{\hat{i}}$ can further be decomposed as $\delta u_{\hat{i}}=\partial_{\hat{i}}\,U+U_{\hat{i}}$. For ideal fluid perturbations, the anisotropic tensor vanishes; therefore, we have not considered it in fluid perturbations. Additionally, $U_{\hat{i}}$ is solenoidal.

\subsection*{Gauge Fixing and Gauge invariant variables}
    
We begin by noting that the teleparallel geometry under consideration remains invariant, up to linear order, under a small perturbation of a cosmologically symmetric background geometry when subjected to an infinitesimal \sout{coordinate transformation} \skmn{diffeomorphism} of the form
\begin{equation}
x'^\mu = x^\mu + \zeta^\mu(x)\label{eq_56}
\end{equation}
as long as the components of the vector field $\zeta^\mu$ are sufficiently small so that the perturbative expansion of the tetrad remains valid. Under such a transformation, the tetrad changes according to the Lie derivative \cite{Bruni:1996im, Malik:2008im}
\begin{equation}
e'^a\,_\mu = e^a\,_\mu + (\mathcal{L}_\zeta \mathring{e})^a\,_\mu\,,\label{eq_57}
\end{equation}
where we retain only linear terms in both the tetrad perturbation and the vector field $\zeta^\mu$. Thus, the background tetrad $\mathring{e}^a{}_\mu$ appears in the second term.

Assuming the transformed tetrad can also be written as a perturbation around the same background,
\begin{equation}
e'^a{}_\mu = \mathring{e}^a{}_\mu + \delta e'^a{}_\mu\,,\label{eq_58}
\end{equation}
We find that the tetrad perturbation transforms as
\begin{equation}
\delta_{\mathring{\zeta}} \delta e^a\,_\mu = \delta e'^a\,_\mu - \delta e^a\,_\mu = (\mathcal{L}_\zeta \mathring{e})^a\,_\mu = \zeta^\nu \partial_\nu \mathring{e}^a\,_\mu + \partial_\mu \zeta^\nu \mathring{e}^a\,_\nu\,.\label{eq_59}
\end{equation}

Similarly, the EM-tensor perturbation transforms as 
\begin{equation}
    \delta_{\mathring{\zeta}} \delta \Theta_{\mu\nu} = (\mathcal{L}_{\zeta} \mathring{\Theta})_{\mu\nu} = \zeta^{\lambda} \partial_\lambda \mathring{\Theta}_{\mu\nu} + (\partial_\nu \zeta^{\lambda}) \mathring{\Theta}_{\mu\lambda} + (\partial_\mu \zeta^{\lambda}) \mathring{\Theta}_{\lambda\nu}\,.\label{eq_60}
\end{equation}
Decomposing the transformation vector $\zeta$ as $[\zeta^\mu]=\left[\zeta^0,\,a^{-1}\delta^{\hat{i}\hat{j}}(\partial_{\hat{j}}\,\zeta+\zeta_{\hat{j}})\right]$, where $\zeta^0$ and $\zeta$ are scalars and $\zeta^{\hat{i}}$ is a divergenceless vector. The perturbations over FLRW transform as:

\begin{subequations}
\label{eq_61}
\begin{align}
    \sigma' &= \sigma\,, \\
     \phi' &=  \phi + \dot{\zeta}^0\,, \\
    \psi' &= \psi + H \zeta^0\,, &
    \sigma^{\hat{k}}{}' &= \sigma^{\hat{k}} + \frac{1}{2a} \epsilon^{\hat{i}\hat{j}\hat{k}} \partial_{\hat{i}} \zeta_{\hat{j}}\,, \\
    \beta' &= \beta + \frac{1}{a} \zeta^0\,, &
    \beta_{\hat{i}}' &= \beta_{\hat{i}}\,, \\
    B' &= B - H \zeta + \dot{\zeta}\,, &
    B_{\hat{i}}' &= B_{\hat{i}} - H \zeta_{\hat{i}} + \dot{\zeta}_{\hat{i}}\,, \\
    E' &= E + \frac{1}{a} \zeta\,, &
    h_{\hat{i}}' &= h_{\hat{i}} + \frac{1}{2a} \zeta_{\hat{i}}\,, &
    h_{\hat{i}\hat{j}}' &= h_{\hat{i}\hat{j}}\,.
\end{align}
\end{subequations}
Whereas the fluid perturbations transform as:
\begin{subequations}
\label{eq_62}
\begin{align}
    \delta \rho' &= \delta \rho + \dot{\rho}\, \zeta^{0}\,, \\
    \delta p' &= \delta p + \dot{p}\, \zeta^{0}\,, \\
    U' &= U + H \zeta - \dot{\zeta}\,, \\
    U_{\hat{i}}' &= U_{\hat{i}} + H \zeta_{\hat{i}} - \dot{\zeta}_{\hat{i}}\,.
\end{align}
\end{subequations}

The irreducible parts of the tetrad and the energy-momentum tensor perturbations transform under infinitesimal \sout{coordinate transformations} \skmn{diffeomorphisms}, so their values depend on the coordinate system. To extract degrees of freedom independent of diffeomorphism-induced non-physicality, two \skmn{\sout{equivalent}} methods are used: gauge fixing and constructing gauge-invariant variables \cite{bardeen1980gauge}. Gauge fixing chooses a reference coordinate system by imposing conditions on certain components (e.g., setting them to zero), treating the remaining components as physical. Alternatively, gauge-invariant variables are formed as linear combinations of perturbations whose gauge shifts cancel.

Any irreducible perturbation component $\hat{Q}$ can be expressed as \cite{Mukhanov:1990me,heisenberg2023gaugeinvariantcosmologicalperturbationsgeneral}
\begin{equation}
\hat{Q} = \hat{Q}_{\text{fixed}} - \delta_{\zeta} \hat{Q}\,,
\label{eq_63}
\end{equation}
where $\hat{Q}_{\text{fixed}}$ is the value in a fixed coordinate system, and $\delta_{\zeta} \hat{Q}$ encodes the change due to the coordinate shift generated by $\zeta$. Under a transformation generated by $\zeta'$, only $\delta_{\zeta} \hat{Q}$ changes, while $\hat{Q}_{\text{fixed}}$ remains invariant. Imposing gauge-fixing conditions on a chosen subset $\{\hat{Q}_i\}$ 
\begin{equation}
0 = \hat{Q}_{i, \text{fixed}}=\hat{Q}_i  +\delta_{\zeta} \hat{Q}_i\,,
\label{eq_64}
\end{equation}
allows us to solve for $\zeta$ in terms of $\hat{Q}_i$. Applying the inverse transformation on other components $\hat{Q}_k$ yields
\begin{equation}
\hat{Q}_{k, \text{fixed}} = \hat{Q}_k + \delta_{\zeta} \hat{Q}_k\,,
\label{eq_65}
\end{equation}
which represents a gauge-invariant linear combination of $\hat{Q}_k$ and $\hat{Q}_i$. An obvious gauge-invariant combination is $B+U$, since their gauge shifts cancel. Additionally pseudoscalar $\sigma$, vector perturbation $\beta_{\hat{i}}$ and tensor perturbations $h_{\hat{i}\hat{j}}$ are manifestly gauge invariant.

In our convention, the Fourier transform of a perturbation $\hat{Q}$ is defined by the following equation, which is used consistently to express cosmological perturbations in Fourier space
\begin{equation}
\hat{Q}(t, \mathbf{x}) = \int \frac{d^3 k}{(2\pi)^{3/2}} \left( \hat{Q}(t, \mathbf{k}) e^{i \mathbf{k} \cdot \mathbf{x}} + \hat{Q}^\dagger (t, \mathbf{k}) e^{-i \mathbf{k} \cdot \mathbf{x}} \right).
\label{eq_66}
\end{equation}

\skmn{Finally, we point out that even though the tetrad perturbations transform both under diffeomorphisms and local Lorentz transformations acting on the Latin indices, the latter do not change the diffeomorphism-invariant observables. Such transformations are exactly compensated by the purely inertial spin connection, which ensures that the Lorentz sector represents pure gauge degrees of freedom governed by the antisymmetric part of the field equations.}

\section{\texorpdfstring{$F(T,T_G)$}{} Cosmological Perturbations} \label{sec:f_T_TG_pert}

In what follows, we will deal with different kinds of perturbations separately.  Components of linear perturbations of teleparallel quantities separately for tensor, (pseudo)vector, and (pseudo)scalar modes are given in Appendix  \ref{app_A}.

\subsection{Tensor perturbations}\label{ten_pert}

As is clear from \eqref{eq_61}, linear tensor perturbations are gauge invariant by construction. This fact does not change in any redefinition of the time coordinate  (for example, in a conformal time description) as it is a direct consequence of the Stewart-Walker lemma \cite{Stewart:1974}. Considering the transverse traceless part of the cosmological perturbations in the tetrad in Eq. \eqref{eq_52} which are 
\begin{equation}
\displaystyle\underset{Tensor}{\delta \,e^{\ell}\,_{\mu}}=\frac{a}{2} \delta^{\hat{i}}\,_{\mu}\delta^{\ell\hat{j}} \,h_{\hat{i}\hat{j}}\label{eq_67} 
\end{equation}

Explicitely traceless and transverse conditions read as   $   h^{\hat{i}}_{\ \hat{i}} = 0$ and $\partial^{\hat{i}} h_{\hat{i}\hat{j}} = 0$ respectively. 

The antisymmetric part of equations, i.e. \eqref{eq_33}, is satisfied identically, whereas the time-time and time-space components of the symmetric part vanish. The space-space symmetric part on shell leads to the gravitational wave propagation equation
\begin{equation}
\ddot{h}_{\hat{i}\hat{j}}+\left(3H+\frac{-\dot{F_T}+4\dot{H}\dot{F_{T_G}}+4H\ddot{F_{T_G}}}{-F_T+4H\dot{F_{T_G}}}\right)\dot{h}_{\hat{i}\hat{j}}-\frac{1}{a^{2}}\triangle h_{\hat{i}\hat{j}}=0\,\label{eq_68}
\end{equation}
which confirms the result provided in \cite{Mishra:2025}. Here $\triangle=\partial^{\hat{i}}\partial_{\hat{i}} \equiv-k^2$ is the spatial Euclidean Laplacian. The result reduces to $F(T)$ case \cite{Cai:2018rzd}, when $F_{T_G}=0$ as expected. From Eq. \eqref{eq_68}, the stability condition $-F_T+4H\dot{F_{T_G}}>0$ can be read, which is the same as that obtained in \cite{Mishra:2025} to avoid ghosts. This is quite similar to standard gravitational wave propagation, except that only the unobservable effect of Hubble friction is altered. Still, the modified Hubble friction term can be constrained using CMB B-modes, by considering the alpha parameterizations described in \cite{Mishra:2025}. Another important aspect is that Eq. \eqref{eq_68} characterizes a massless spin-2 field propagating at the speed of light. Consequently, the propagation of gravitational waves aligns fully with the observations from the multimessenger event GW170817 \cite{TheLIGOScientific:2017qsa} and the corresponding gamma-ray burst  \cite{Goldstein:2017mmi}.

\subsection{Vector and pseudovector perturbations}\label{vec_pert}

Further breaking down cosmological perturbations involves considering vector and pseudovector perturbations, which are solenoidal. Typically, in an expanding universe, such vector modes decay unless maintained by anisotropic stress. Although we do not expect a different outcome here, it is still essential to examine these cases carefully to be sure.  
 The vector and pseudovector parts of perturbations from \eqref{eq_52} are given by  

\begin{equation}
    \underset{Vector}{\left[\delta e^{\ell}\,_{\mu}\right]} = \left[\begin{array}{cc}
    0 & a\beta_{\hat{i}} \\
   B^{\tilde{l}} & a\delta^{\tilde{l}\,\hat{i}}\left( 2\partial_{(\hat{i}}h_{\hat{j})}+\epsilon_{\hat{i}\hat{j}\hat{k}}\sigma^{\hat{k}}\right)
    \end{array}\right] \,.\label{eq_69}
\end{equation}
Using the Eq. \eqref{eq_33}, we see that mixed ($0i$) and spatial ($ij$) components of antisymmetric parts of equations are:

\begin{equation}
\underset{Vector}{\mathbb{E}_{[0\,\hat{i}]}} \equiv
\left( \dot{F_T} + 4H^2 \dot{F_{T_G}} \right)
\left( \triangle h_{\hat{i}} - \epsilon_{\hat{i}\hat{j}\hat{k}}\, \partial^{\hat{j}}\sigma^{\hat{k}} \right)
+ \frac{2H}{a}\, \dot{F_{T_G}}\, \triangle \left( \beta_{\hat{i}} - B_{\hat{i}} + 2 a \dot{h}_{\hat {i}} \right)
= 0\,,\label{eq_70}
\end{equation}

\begin{equation}
\underset{Vector}{\mathbb{E}_{[\hat{i}\,\hat{j}]}} \equiv\left( \dot{F_T} + 4H^2 \dot{F_{T_G}} \right) \left( \partial_{\hat{i}} \beta_{\hat{j}} - \partial_{\hat{j}} \beta_{\hat{i}} \right) = 0\,.
\label{eq_71}
\end{equation}

After taking divergence, \eqref{eq_71} implies
\begin{equation}
\left( \dot{F_T} + 4H^2 \dot{F_{T_G}} \right) \triangle \beta_{\hat{j}} = 0\,,
\label{eq_72}
\end{equation}
which for $ \dot{F_T} + 4H^2 \dot{F_{T_G}}\neq 0$ should be solved as $\beta_{\hat{i}}=0$ in perturbation theory, due to Liouville's theorem. Considering this constraint and since $\beta_{\hat{i}}$ is gauge invariant as well as  defining the gauge invariant quantities,
\begin{subequations}
  \begin{align}
    \mathcal{B}_{\hat{i}} &= B_{\hat{i}} - 2 a \dot{h}_{\hat{i}}\,, \\ 
    \mathcal{U}_{\hat{i}} &= U_{\hat{i}} + B_{\hat{i}} \,,\\
    \mathcal{\beth}_{\hat{i}} &= \sigma_{\hat{i}} + \epsilon_{\hat{i}\hat{j}\hat{k}}\, \partial^{\hat{j}} h^{\hat{k}}\,.
  \end{align}
  \label{eq_73}
\end{subequations}
In terms of the above gauge invariant quantities, \eqref{eq_70} can be written as
\begin{equation}
-\left( \dot{F_T} + 4H^2 \dot{F_{T_G}} \right)\, \epsilon_{\hat{i}\hat{j}\hat{k}}\, \partial^{\hat{j}} \mathcal{\beth}^{\hat{k}}
 - \frac{2H}{a}\, \dot{F_{T_G}}\, \triangle \mathcal{B}_{\hat{i}} = 0\,.
\label{eq_74}
\end{equation}
Symmetric part of Eqs. \eqref{eq_29} are given as,
\begin{eqnarray}
   \underset{Vector}{ \mathbb{E}_{(0\,\hat{i})}} &\equiv&
    \triangle\mathcal{B}_{\hat{i}} \left(-F_T + 2H\,\dot{F_{T_G}}\right)
    \;-\; a \left(\dot{F_T} + 4H^{2}\,\dot{F_{T_G}}\right)
    \epsilon_{\hat{i}\hat{j}\hat{k}}\,\partial^{\hat{j}}\,\mathcal{\beth}^{\hat{k}}
    = 2\kappa^2 a^2\,\mathcal{U}_{\hat{i}}\,(\rho + p)\,,
    \label{eq_75} \\
    \underset{Vector}{\mathbb{E}_{(\hat{i}\,\hat{j})}} &\equiv&
    \left(\partial_{\hat{i}}\dot{\mathcal{B}}_{\hat{j}} + \partial_{\hat{j}}\dot{\mathcal{B}}_{\hat{i}}\right)
    + \left(\partial_{\hat{i}}\mathcal{B}_{\hat{j}} + \partial_{\hat{j}}\mathcal{B}_{\hat{i}}\right)
    \left(
        2H + \frac{ -\dot{F_T} + 4\dot{H}\dot{F_{T_G}} + 4H\ddot{F_{T_G}} }
                    { -F_T + 4H\dot{F_{T_G}} }
    \right) = 0\,.
    \label{eq_76}
\end{eqnarray}
 Under antisymmetric part \eqref{eq_74}, Eq. \eqref{eq_75} reduces to:
 \begin{equation}
      \triangle\mathcal{B}_{\hat{i}} \left(-F_T + 4H\,\dot{F_{T_G}}\right) = 2\kappa^2 a^2\,\mathcal{U}_{\hat{i}}\,(\rho + p)\,,\label{eq_77}
 \end{equation}
 where $\rho$ and $p$ can be substituted from the background equations. \eqref{eq_39} and \eqref{eq_40}.
Additionally, Eq. \eqref{eq_76} can be written as,
\begin{equation}
    \dot{\mathcal{B}}
  +
    \left(
        2H + \frac{ -\dot{F_T} + 4\dot{H}\dot{F_{T_G}} + 4H\ddot{F_{T_G}} }
                    { -F_T + 4H\dot{F_{T_G}} }
    \right)\mathcal{B} = 0\,.\label{eq_78}
\end{equation}
Eqs. \eqref{eq_77} and \eqref{eq_78} involve only two components, the gauge invariant geometric perturbation $\mathcal{B}_{\hat{i}}$ and the vortical part of fluid component encoded in the gauge invariant combination $\mathcal{U}_{\hat{i}}$. We immediately see that \eqref{eq_78} is a constraint equation and \eqref{eq_77} is a Poisson-like equation. Therefore, it is clear that vector modes are not propagating. Furthermore, it is obvious that if we can solve the system for $\mathcal{B}_{\hat{i}}$, then one can solve it for $\mathcal{U}_{\hat{i}}$ as well. Additionally, one can read the stability condition  $-F_T+4H\dot{F_{T_G}}>0$ off from \eqref{eq_78}, which is what was obtained in \cite{Mishra:2025} and realized in \ref{ten_pert}. We also see that the contribution from the pseudovector is constrained by the antisymmetric part of the equations, which makes perfect sense given that the pseudovector is essentially a Lorentz variable. Thus, we get a well-behaved vector sector in the theory.

\subsection{Scalar and pseudoscalar perturbations}
We now move to the most important sector of cosmological perturbations: the scalar part. These represent fluctuations in density and pressure that seed the formation of large-scale structures such as galaxies and clusters. Also, these are the primary sources of the temperature anisotropies observed in the cosmic microwave background (CMB), providing a direct link to the physics of the early universe. Unlike vector modes, which decay, and tensor modes, which are subdominant, scalar modes grow under gravitational instability. The scalar part of perturbations is obtained by switching off all the vector and tensor modes from the perturbed tetrad as,
\begin{equation}
   \underset{Scalar}{\left[\delta\, e^{a}\,_{\mu}\right]}  = \left[\begin{array}{cc}
     \phi & a\partial_{\hat{i}}\beta \\
    \delta^{\tilde{l}}{}_{\hat{i}}\partial^{\hat{i}}B & a\delta^{\tilde{l}\,\hat{i}}\left(\psi\delta_{\hat{i}\hat{j}}+\partial_{\hat{i}}\partial_{\hat{j}}E+\epsilon_{\hat{i}\hat{j}\hat{k}}\partial^{\hat{k}}\sigma\right)
    \end{array}\right]. \label{eq_79}
\end{equation}
 All the components of the antisymmetric part  Eq. \eqref{eq_33} vanish except for mixed components, for which we get 
\begin{equation}
\underset{Scalar}{\mathbb{E}_{[0\,\hat{i}]}} \equiv 
\partial_{\hat{i}}\left(H \delta F_T - \psi \dot{F_T} + 4 H \dot{F_{T_G}} \left( H( \phi - \psi) - \dot{\psi} \right) + \frac{T_G \delta F_{T_G}}{6 H}\right) = 0\,.
\label{eq_80}
\end{equation}
While the components of the symmetric part of equations (on shell) are given by,
\begin{align}
&\underset{Scalar}{\mathbb{E}_{00}} \equiv 
-\frac{2 k^2 \big[
    F_T \big(\psi + a H (B - \beta - a \dot{E}) \big) 
    + 2 a H^2 \dot{F_{T_G}} \big(\beta - 3 B + 3 a \dot{E} \big) 
    - 2 H^2 \delta F_{T_G}
\big]}{a^2}
- \frac{1}{2} T_G \delta F_{T_G}
- 6 H^2 \delta F_T \nonumber\\[1.0ex]
&\quad 
+ 6 H \big[
    2 H \dot{F_{T_G}} (3 \dot{\psi} - 4 H  \phi) 
    + F_T (H  \phi - \dot{\psi}) 
    + 2 H^2 \dot{\delta F_{T_G}}
\big] 
= \kappa^2 \delta \rho\,,
\label{eq_81}\\[2.0ex]
&\int_{\hat{i}}\underset{Scalar}{\mathbb{E}_{(0\,\hat{i})}} \equiv 
4 H \dot{F_{T_G}} \big(\dot{\psi} - H (2  \phi + \psi)\big) 
+ 2 F_T \big(H \phi - \dot{\psi}\big) 
- \psi \dot{F_T} 
+ 4 H^2 \dot{\delta F_{T_G}} 
+ 4 \dot{H} H \delta F_{T_G} 
- H \delta F_T 
\nonumber\\
&\quad
= a \kappa^2 (P + \rho) (B - \beta + U)\,,
\label{eq_82}\\[2.0ex]
&\int_{\hat{i}\hat{j}} \underset{Scalar}{\mathbb{E}_{\hat{i}\hat{j}}} \equiv  
a^2 \big[
    4 \big(H \ddot{E} + \dot{E} (3 H^2 + \dot{H})\big) \dot{F_{T_G}}
    + 4 \dot{E} H \ddot{F_{T_G}}
    - F_T (\ddot{E} + 3 \dot{E} H)
    - \dot{E} \dot{F_T}
\big] \nonumber\\
&\quad 
+ a \big[
    -4 B H \ddot{F_{T_G}}
    - 4 \big(H^2 (2 B - \beta) + \dot{B} H + B \dot{H}\big) \dot{F_{T_G}}
    + 2 B H F_T + B \dot{F_T} + \dot{B} F_T - 2 \beta  H F_T - \dot{\beta} F_T
\big] \nonumber\\
&\quad
+ \psi \big(F_T - 4 H \dot{F_{T_G}}\big) + \phi  F_T
= 0\,,
\label{eq_83}\\[2.0ex]
&\underset{Scalar}{\mathbb{E}^{\hat{i}}\,_{\hat{i}}} \equiv  
F_T \Bigg[
    \frac{
        k^2 \big[
            2 a \big(
                a (\ddot{E} + 3 \dot{E} H) + \dot{\beta} - \dot{B}
            \big)
            - 2 \big(2 a H (B - \beta) +  \phi + \psi \big)
        \big]
    }{3 a^2}
    - 2 \ddot{\psi}
    + 2 H (3 H  \phi + \dot{\phi} - 3 \dot{\psi})
    + 4 \dot{H}  \phi
\Bigg] \nonumber\\
&\quad
+ \dot{F_{T_G}} \Bigg[
    -\frac{8 k^2 \big[
        H \big(
            a^2 (\ddot{E} + 3 \dot{E} H)
            + a H (\beta - 2 B) 
            - \psi 
        \big)
        + a \dot{H} (a \dot{E} - B)
        - a \dot{B} H
    \big]}{3 a^2}
    + 8 H \ddot{\psi} 
    - \frac{4  \phi T_G}{3 H}
    - 12 H^2 \dot{\phi} \nonumber \\ 
&\quad
+ 8 (3 H^2 + \dot{H}) \dot{\psi}
\Bigg]+ \frac{8 H k^2 (B - a \dot{E}) \ddot{F_{T_G}}}{3a}
+ 8 H (\dot{\psi} - 2 H  \phi) \ddot{F_{T_G}}
+ 4 H^2 \ddot{\delta F_{T_G}}
+ \dot{F_T} \left[
    \frac{2}{3} k^2 \Big(\dot{E} - \frac{B}{a}\Big) 
    + 4 H  \phi - 2 \dot{\psi} 
\right] \nonumber\\
&\quad
+ 8 (H^2 + \dot{H}) H \dot{\delta F_{T_G}}
- \frac{1}{2} T_G \delta F_{T_G}
- 2 (3 H^2 + \dot{H}) \delta F_T
- 2 H \dot{\delta F_T}
= -\kappa^2 \delta p\,,
\label{eq_84}
\end{align}
where $\delta F_T=F_{TT}\,\delta T+F_{TT_G}\,\delta T_G$ ,  $\delta F_{T_G}=F_{TT_G}\,\delta T+F_{T_GT_G}\,\delta T_G$ and the equations are expressed in Fourier domain. In these expressions, $T_G= 24 H^{2} \left(\dot{H} + H^{2}\right)$ is background teleparallel Gauss-Bonnet term. The linearized Eqs. \eqref{eq_80}-\eqref{eq_84} are in gauge-ready form, that is, being most general, they are free from any gauge choices. An important point to note is that the pseudoscalar does not appear in the field equations. It can therefore be morphed as a remnant symmetry. In what follows, we shall write these in gauge-invariant form and also specify various gauge choices that are useful for different purposes. 

Consider the following gauge-invariant combinations of geometrical perturbations,
\begin{subequations}
  \begin{align}
    \mathcal{X}_1 &=  \phi-\psi-a\dot{\beta}\,, \\ 
     \mathcal{X}_2 &=\psi-\dot{a}\beta \,,\\
    \mathcal{X}_3 &= -B+a\dot{E}\,,
  \end{align}
  and the gauge-invariant matter perturbations

  \begin{align}
    \mathcal{\delta\varrho} &= \delta\rho-a\dot{\rho}\beta\,, \\ 
 \mathcal{\delta P} &= \delta p-a\dot{p}\beta\,, \\
    \mathcal{U}_s &= B+U\,.
  \end{align}
  \label{eq_85}
\end{subequations}
In terms of these gauge invariant combinations, the linearized equations take the form,
\begin{align}
    \underset{GI}{\mathbb{E}_{00}}&=\kappa^2  \mathcal{\delta \varrho}\,,\label{eq_86}\\
    \int_{\hat{i}}\underset{GI}{\mathbb{E}_{(0\,\hat{i})}} &=a\kappa^2  \mathcal{U}_s\,,\label{eq_87}\\
    \int_{\hat{i}\hat{j}} \underset{GI}{\mathbb{E}_{\hat{i}\hat{j}}}&=0\,,\label{eq_88}\\
    \underset{GI}{\mathbb{E}^{\hat{i}}\,_{\hat{i}}}&=-\kappa^2 \mathcal{\delta P}\,,\label{eq_89}
\end{align}
and the anti-symmetric part
\begin{align}
  \int_{\hat{i}}\underset{GI}{\mathbb{E}_{[0\,\hat{i}]}}&=0\,.\label{eq_90}   
\end{align}
Full expressions of \eqref{eq_86}-\eqref{eq_90}  in terms of gauge invariant variables are given in \cite{Shivamgit2025}. For demonstration, we present the anisotropic part \eqref{eq_88} here
\begin{equation}
a \Bigg[
    \mathcal{X}_3 \left(
        4 H \ddot{F_{T_G}} + 8 H^2 \dot{F_{T_G}}
        + 4 \dot{H} \dot{F_{T_G}} - 2 H F_T - \dot{F_T}
    \right)
    - \dot{\mathcal{X}}_3 \left(F_T - 4 H \dot{F_{T_G}}\right)
\Bigg]
+ 2 \mathcal{X}_2 \left(F_T - 2 H \dot{F_{T_G}}\right)
+ \mathcal{X}_1 F_T=0\,.\label{eq_91}
\end{equation}

An important point to note here is that all these equations are on shell, i.e., the span of all background contributions \set{\eqref{eq_39},\eqref{eq_40},\eqref{eq_41}} is used to obtain the equations in the form presented.

\section{Gauge Choices}\label{sec:gauge_ch}

Since tensor and pseudoscalar perturbations are manifestly gauge invariant, no gauge choices affect them. Also, as demonstrated in Sec. \ref{vec_pert}, the vector perturbations decay, so gauge choices are not much of interest in these sectors as well. We therefore proceed with gauge choices of scalar perturbations as follows: 
\subsection{Newtonian gauge}\label{newton_gauge}
The Newtonian / longitudinal gauge is widely studied in teleparallel gravity \cite{Wu_2012, Zheng_2011, Bahamonde:2020lsm}. This gauge is particularly important for calculating the effective gravitational constant in modified theories, and the remaining modes correspond to Bardeen potentials \skmn{in longitudinal gauge} after using the antisymmetric parts of the equations. This choice is framed as $E=0$ and $B=\beta$, transforming the perturbed metric into diagonal form
\begin{equation}
\underset{NG}{\left[\delta g_{\mu\nu}\right]} = \left[\begin{array}{cc}
-2 \phi & 0\\
0 & 2a^{2}\psi\delta_{\hat{i}\hat{j}}
\end{array}\right]\,. \label{eq_92}
\end{equation}
We present here equations in Newtonian gauge after solving for $\delta F_T$ from the antisymmetric part of equations.
\begin{align}
&\underset{NG}{\mathbb{E}_{00}} \equiv 
F_T \left[ 6 H (H  \phi - \dot{\psi}) - \frac{2 k^2 \psi}{a^2} \right]
+ \delta F_{T_G} \left( \frac{4 H^2 k^2}{a^2} + \frac{T_G}{2} \right)
- 6 H \psi \dot{F_T}
+ 12 H^3 \dot{\delta F_{T_G}} \nonumber\\
&\quad
+ \dot{F_{T_G}} \left[ 
    \frac{8 \beta H^2 k^2}{a} 
    - 12 H^2 \big( 2 H ( \phi + \psi) - \dot{\psi} \big) 
\right]
= \kappa^2 \delta \rho\,,
\label{eq_93}\\[2.0ex]
&\int_{\hat{i}}\underset{NG}{\mathbb{E}_{0\,\hat{i}}} \equiv 
2 \psi \dot{F_T} 
- 4 H^2 \dot{\delta F_{T_G}}
+ 4 H^2 ( \phi + 2 \psi) \dot{F_{T_G}} 
+ F_T \big( 2 \dot{\psi} - 2 H  \phi \big)- 4 (H^2 + 2 \dot{H}) H \delta F_{T_G} \nonumber\\
&\hspace{12cm}
= - a \kappa^2 U (p + \rho)\,,
\label{eq_94}\\[2.0ex]
&\int_{\hat{i}\hat{j}}\underset{NG}{\mathbb{E}_{\hat{i}\hat{j}}} \equiv 
\beta \dot{F_T}
+ \frac{F_T ( \phi + \psi)}{a} 
- \frac{4}{a} \Big[
    a \beta H \ddot{F_{T_G}} 
    + \dot{F_{T_G}} \big( H (a \dot{\beta} + a \beta H + \psi) + a \beta \dot{H} \big)
\Big] 
= 0\,,
\label{eq_95}\\[2.0ex]
&\underset{NG}{\mathbb{E}^{\hat{i}}\,_{\hat{i}}} \equiv
4 H \Bigg[
    H \big( \ddot{\delta F_{T_G}} - 2 ( \phi + \psi) \ddot{F_{T_G}} \big)
    + \frac{T_G \dot{\delta F_{T_G}}}{6 H^2}
\Bigg]
+ F_T \big[
    -2 \ddot{\psi} + 2 H (3 H  \phi + \dot{\phi} - 3 \dot{\psi}) + 4 \dot{H}  \phi
\big] \nonumber\\
&\quad
+ \dot{F_T} \big( 4 H  \phi - 6 H \psi - 4 \dot{\psi} \big)
- 2 \psi \ddot{F_T}
+ 4 \big( 2 H \ddot{H} + 3 H^4 + 9 \dot{H} H^2 + 2 \dot{H}^2 \big) \delta F_{T_G} \nonumber\\
&\quad
- 4 H \dot{F_{T_G}} \Big[
    4 \dot{H} ( \phi + \psi) 
    + H \big( 2 ( H \phi + 3 H \psi + \dot{\psi} ) + \dot{\phi} \big)
\Big]
= - \kappa^2 \delta p\,.
\label{eq_96}
`\end{align}

Though we have presented the symmetric equations after inserting antisymmetric constraint as eliminating $\delta F_T$ and its derivatives, for calculation of effective gravitational constant or lensing parameter, one must take the following path: Solve antisymmetric part of equations for Lorentz variable $\beta$ in terms of potentials $ \phi$ and $\psi$, use symmetric part of equations (usually spatial anisotropic part, mixed and time-time components) and the perturbed matter conservation equation explicitly given in Appendix  \ref{app_B} in their sub-horizon limit and calculate required parameters using standard methods.

\subsection{Synchronous gauge}\label{sync_gauge}
As the name suggests, synchronous gauge is one where the time coordinate matches the proper time of co-moving observers. It begins from a constant-time hypersurface with space-like intervals, and geodesics normal to this surface define the time direction. It is defined by $ \phi \,=0$ and  $B\,=\,\beta$. The choice of hypersurface is not unique and allows any time-independent transformation of the spatial coordinates. The choice of coordinates for fixing the gauge, therefore, is not unique. Hence, it needs additional gauge fixing, which is done for the problem at hand. However, this gauge is widely used in numerical studies and is the primary choice in Boltzmann solvers such as CLASS and CAMB. The tetrad perturbation in this gauge takes the form
\begin{equation}
   \underset{SG}{\left[\delta\, e^{a}\,_{\mu}\right]}  = \left[\begin{array}{cc}
    0 & a\partial_{\hat{i}}\beta \\
    \delta^{\tilde{l}}{}_{\hat{i}}\partial^{\hat{i}}\beta & a\delta^{\tilde{l}\,\hat{i}}\left(\psi\delta_{\hat{i}\hat{j}}+\partial_{\hat{i}}\partial_{\hat{j}}E\right)
    \end{array}\right]\,. \label{eq_97}
\end{equation}
The linearized equations in the synchronous gauge are given below, 
where using the antisymmetric part \eqref{eq_80} we have replaced $\delta F_T$ and its derivatives.
\begin{align}
&\underset{SG}{\mathbb{E}_{00}} \equiv 
F_T \Big[ k^2 \big( 2 \dot{E} H - \frac{2 \psi}{a^2} \big) - 6 H \dot{\psi} \Big]
+ \delta F_{T_G} \left( \frac{4 H^2 k^2}{a^2} + \frac{T_G}{2} \right)
- 6 H \psi \dot{F_T}
+ 12 H^3 \dot{\delta F_{T_G}} \nonumber\\
&\quad
+ \dot{F_{T_G}} \Big[ 
    \frac{4 H^2 k^2 (2 \beta - 3 a \dot{E})}{a} 
    + 12 H^2 (\dot{\psi} - 2 H \psi) 
\Big]
= \kappa^2 \delta \rho\,,
\label{eq_98}\\[2.0ex]
&\int_{\hat{i}}\underset{SG}{\mathbb{E}_{0\,\hat{i}}} \equiv
8 H^2 \psi \dot{F_{T_G}}
+ 2 \psi \dot{F_T} 
+ 2 \dot{\psi} F_T 
- 4 H^2 \dot{\delta F_{T_G}}
- 4 (H^2 + 2 \dot{H}) H \delta F_{T_G}
= - a \kappa^2 U (p + \rho)\,,
\label{eq_99}\\[2.0ex]
&\int_{\hat{i}\hat{j}} \underset{SG}{\mathbb{E}_{\hat{i}\hat{j}}} \equiv
\dot{F_{T_G}} \Big[
    4 H \big( \dot{\beta} - a ( \ddot{E} + 3 \dot{E} H ) \big)
    + \beta H
    + \frac{\psi}{a}
    + 4 \dot{H} (\beta - a \dot{E})
\Big]+ 4 H (\beta - a \dot{E}) \ddot{F_{T_G}}+ F_T \left( a \ddot{E} + 3 a \dot{E} H - \frac{\psi}{a} \right) \nonumber\\
&\quad
- \dot{F_T} (\beta - a \dot{E})
= 0\,,
\label{eq_100}\\[2.0ex]
&\underset{SG}{\mathbb{E}^{\hat{i}}\,_{\hat{i}}} \equiv
\dot{F_{T_G}} \Bigg[
    \frac{8 k^2}{3 a^2} \Big(
        H \big( -a^2 \ddot{E} - 3 a^2 \dot{E} H + a \dot{\beta} + a \beta H + \psi \big)
        + a \dot{H} (\beta - a \dot{E})
    \Big)
    - 8 H \big( 2 \dot{H} \psi + H (3 H \psi + \dot{\psi}) \big)
\Bigg] \nonumber\\
&\quad
+ F_T \Bigg[
    \frac{2}{3} k^2 \left( \ddot{E} - \frac{\psi}{a^2} + 3 \dot{E} H \right)
    - 2 \big( \ddot{\psi} + 3 H \dot{\psi} \big)
\Bigg]
- \dot{F_T} \left(
    \frac{2 k^2 (\beta - a \dot{E})}{3 a} + 6 H \psi + 4 \dot{\psi}
\right)+ \frac{2 T_G \dot{\delta F_{T_G}}}{3 H} \nonumber\\
&\quad
+ \frac{8 H k^2 (\beta - a \dot{E}) \ddot{F_{T_G}}}{3 a} 
- 8 H^2 \psi \ddot{F_{T_G}} 
- 2 \psi \ddot{F_T} 
+ 4 H^2 \ddot{\delta F_{T_G}} + 4 \big( 2 H \ddot{H} + 3 H^4 + 9 \dot{H} H^2 + 2 \dot{H}^2 \big) \delta F_{T_G}
= - \kappa^2 \delta p\,.
\label{eq_101}
\end{align}
Background density $\rho$ and pressure $p$ in \eqref{eq_99} can be eliminated using background equations. Also, there is no symmetrisation symbol on indices in \eqref{eq_99} since substituting the antisymmetric part into the mixed part of the equations renders both $\hat{i}\,0$ and $0\,\hat{i}$ equal.

\subsection{Flat gauge}\label{flat_gauge}

The (Spatially) flat gauge is obtained by setting  $\psi=0$ and $E=0$, rendering projected metric on spatial hypersurfaces unperturbed. This is particularly useful in early universe especially for calculating observables during inflation. The perturbation of tetrad in this gauge is realized as
\begin{equation}
   \underset{FG}{\left[\delta\, e^{a}\,_{\mu}\right]}  = \left[\begin{array}{cc}
     \phi & a\partial_{\hat{i}}\beta \\
    \delta^{\tilde{l}}{}_{\hat{i}}\partial^{\hat{i}}B & 0
    \end{array}\right]\,, \label{eq_102}
\end{equation} and the metric perturbation takes the form 
\begin{equation}
\underset{FG}{\left[\delta g_{\mu\nu}\right]} = \left[\begin{array}{cc}
-2 \phi & a\,\partial_{\hat{i}}(B-\beta)\\
a\,\partial_{\hat{i}}(B-\beta) & 0
\end{array}\right]. \label{eq_103}
\end{equation}
The symmetric field equations after elimination of $\delta F_T$ from antisymmetric constraint are: 
\begin{align}
&\underset{FG}{\mathbb{E}_{00}} \equiv
\delta F_{T_G} \left( \frac{4 H^2 k^2}{a^2} + \frac{T_G}{2} \right)
+ \dot{F_{T_G}} \left[ \frac{4 H^2 k^2 (3 B - \beta)}{a} - 24 H^3 \phi \right]
+ F_T \left[ \frac{2 H k^2 (\beta - B)}{a} + 6 H^2 \phi \right]\nonumber\\
&\hspace{13cm}+ 12 H^3 \dot{\delta F_{T_G}} = \kappa^2 \delta \rho\,,
\label{eq_104}\\[2.0ex]
&\int_{\hat{i}}\underset{FG}{\mathbb{E}_{0\,\hat{i}}} \equiv
8 H^2 \phi \dot{F_{T_G}}
- 4 H \phi F_T
- 8 H^2 \dot{\delta F_{T_G}}
- 8 (H^2 + 2 \dot{H}) H \delta F_{T_G}
= - a \kappa^2 (p + \rho) (B - \beta + U)\,,
\label{eq_105}\\[2.0ex]
&\int_{\hat{i}\hat{j}} \underset{FG}{\mathbb{E}_{\hat{i}\hat{j}}} \equiv
4 B H \ddot{F_{T_G}}
+ F_T \left( -\frac{\phi}{a} + \dot{\beta} - 2 B H - \dot{B} + 2 \beta H \right)
+ \left( 4 B (2 H^2 + \dot{H}) + 4 \dot{B} H - 4 \beta H^2 \right) \dot{F_{T_G}}
- B \dot{F_T}
= 0\,,
\label{eq_106}\\[2.0ex]
&\underset{FG}{\mathbb{E}^{\hat{i}}\,_{\hat{i}}} \equiv
\frac{8 B H k^2 \ddot{F_{T_G}}}{3 a} 
+ 4 H^2 \left( \ddot{\delta F_{T_G}} - 2 \phi \ddot{F_{T_G}} \right)
+ 4 \big( 2 H \ddot{H} + 3 H^4 + 9 \dot{H} H^2 + 2 \dot{H}^2 \big) \delta F_{T_G} + \dot{F_T} \left( 4 H \phi - \frac{2 B k^2}{3 a} \right)\nonumber\\
&\quad
+ F_T \Bigg[
    -\frac{2 k^2 \left( -a \dot{\beta} + 2 a B H + a \dot{B} - 2 a \beta H + \phi \right)}{3 a^2}
    + 4 \dot{H} \phi
    + 2 H ( 3 H \phi + \dot{\phi} )
\Bigg] + \frac{2 T_G \dot{\delta F_{T_G}}}{3 H}\nonumber\\
&\quad
+ \dot{F_{T_G}} \Bigg[
    \frac{8 k^2 \left( H^2 (2 B - \beta) + \dot{B} H + B \dot{H} \right)}{3 a}
    - 4 H \big( 4 \dot{H} \phi + H ( 2 H \phi + \dot{\phi} ) \big)
\Bigg]
= - \kappa^2 \delta p\,.
\label{eq_107}
\end{align}
We stress on the fact that in \eqref{eq_105}, $\kappa^2(\rho+p)$ can be obtained by combining \eqref{eq_39} and \eqref{eq_40} and so the product  $\kappa^2(\rho+p)(B-\beta)$  can be transferred to LHS, yielding constraining equation for velocity component $U$. Thereafter the equations can be utilized for numerical studies etc.

\subsection{Comoving gauge}\label{comov_gauge}
All the gauge choices above correspond to geometric part of perturbations. We introduce a gauge arising from freedom of choice of matter perturbations viz. comoving gauge. This is well studied in literature and is widely used in studies related to curvature perturbations. In this choice, both the slicing and the threading are assumed to be comoving \cite{Clifton:2020}, therefore setting the conditions $B\,=\,\beta$ and the velocity perturbation $\delta u_{\hat{i}}=0$ which for scalar perturbations read as $U=0$. The equations can therefore be obtained by substituting above conditions in gauge free equations. For this gauge choice, we explicitly express all components of symmetric and antisymmetric parts of field equations as,

\begin{align}
&\int_{\hat{i}}\underset{CG}{\mathbb{E}_{[0\,\hat{i}]}} \equiv 
H \delta F_T
- \psi \dot{F_T}
+ 4 H \dot{F_{T_G}} \big[ H( \phi - \psi) - \dot{\psi} \big]
+ \frac{T_G \delta F_{T_G}}{6 H}
= 0\,,
\label{eq_108}\\[2.0ex]
&\underset{CG}{\mathbb{E}_{00}} \equiv 
F_T \Big[ k^2 \big( 2 \dot{E} H - \frac{2 \psi}{a^2} \big) + 6 H \big( H \phi - \dot{\psi} \big) \Big]
+ \delta F_{T_G} \left( \frac{4 H^2 k^2}{a^2} + \frac{T_G}{2} \right) \nonumber\\
&\quad
+ \dot{F_{T_G}} \Big[ \frac{4 H^2 k^2 (2 \beta - 3 a \dot{E})}{a}
    - 12 H^2 \big( 2 H ( \phi + \psi) - \dot{\psi} \big) \Big]
- 6 H \psi \dot{F_T}
+ 12 H^3 \dot{\delta F_{T_G}}
= \kappa^2 \delta \rho\,,
\label{eq_109}\\[2.0ex]
&\int_{\hat{i}}\underset{CG}{\mathbb{E}_{(0\,\hat{i})}} \equiv
4 H \dot{F_{T_G}} \big( \dot{\psi} - H (2 \phi + \psi) \big)
+ 2 H \phi F_T
- \psi \dot{F_T}
- 2 \dot{\psi} F_T
+ 4 H^2 \dot{\delta F_{T_G}}
+ 4 \dot{H} H \delta F_{T_G}
- H \delta F_T
= 0\,,
\label{eq_110}\\[2.0ex]
&\int_{\hat{i}\hat{j}} \underset{CG}{\mathbb{E}_{\hat{i}\hat{j}}} \equiv
- \frac{F_T \big( - a^2 ( \ddot{E} + 3 \dot{E} H ) + \phi + \psi \big)}{a}
+ 4 H (\beta - a \dot{E}) \ddot{F_{T_G}}
+ \dot{F_T} (a \dot{E} - \beta) \nonumber\\
&\quad
+ \dot{F_{T_G}} \Big[
    -4 a \big( H \ddot{E} + \dot{E} (3 H^2 + \dot{H}) \big)
    + \frac{4 H \psi}{a}
    + \frac{\beta T_G}{6 H^2}
    + 4 \dot{\beta} H
\Big]
= 0\,,
\label{eq_111}\\[2.0ex]
&\underset{CG}{\mathbb{E}^{\hat{i}}\,_{\hat{i}}} \equiv
\dot{F_{T_G}} \Bigg[
    \frac{k^{2}}{9} \Big(
        -24 ( H \ddot{E} + \dot{E} (3 H^{2} + \dot{H}) )
        + \frac{24 H \psi}{a^{2}}
        + \frac{\beta T_{G}}{a H^{2}}
        + \frac{24 \dot{\beta} H}{a}
    \Big)
    + 8 H \ddot{\psi}
    - \frac{4 \phi T_{G}}{3 H}
    - 12 H^{2} \dot{\phi}
    + 8 (3 H^{2} + \dot{H}) \dot{\psi}
\Bigg] \nonumber\\
&\quad
+ F_{T} \Bigg[
    -\frac{2 k^{2}}{3 a^{2}} \big( - a^{2} (\ddot{E} + 3 \dot{E} H) + \phi + \psi \big)
    - 2 \ddot{\psi}
    + 2 H (3 H \phi + \dot{\phi} - 3 \dot{\psi})
    + 4 \dot{H} \phi
\Bigg]- 2 (3 H^{2} + \dot{H}) \delta F_{T}
- 2 H \dot{\delta F_{T}} \nonumber\\
&\quad
+ \frac{8 H k^{2} (\beta - a \dot{E}) \ddot{F_{T_G}}}{3 a}
+ 8 H (\dot{\psi} - 2 H \phi) \ddot{F_{T_G}}
+ 4 H^{2} \ddot{\delta F_{T_G}}+ \dot{F_{T}} \left[ \frac{2}{3} k^{2} \left( \dot{E} - \frac{\beta}{a} \right) + 4 H \phi - 2 \dot{\psi} \right]
  \nonumber\\
&\quad
+ 8 (H^{2}+ \dot{H}) H \dot{\delta F_{T_{G}}}
- \frac{1}{2} T_{G} \delta F_{T_{G}}
= - \kappa^{2} \delta p\,.
\label{eq_112}
\end{align}

Another matter-gauge choice is the so-called ``total matter gauge'', which we do not discuss in full detail here but end with a concise introduction. It is characterized by a vanishing momentum-density perturbation, which means the total velocity potential vanishes. This gauge is often used to analyze large-scale adiabatic perturbations and is particularly important when dealing with multiple fluid systems. In terms of perturbed quantities, it is characterized by $B-\beta+U=0$ and $E=0$.

\section{Conclusion} \label{sec:conc}

In the presented work, we have developed a gauge-invariant formulation of cosmological perturbations within the $F(T, T_G)$ theory of gravity. Since physical observables must remain independent of coordinate choices, constructing gauge-invariant variables provides a consistent and unambiguous description of perturbations beyond general relativity. Introducing background, we moved to the perturbed field equations, where we identified the combinations of metric and matter perturbations that remain invariant under infinitesimal \sout{coordinate transformations} \skmn{diffeomorphisms} and derived the corresponding evolution equations for scalar, vector, pseudoscalar, pseudovector, and tensor modes. In the appropriate limit, our formalism smoothly recovers the standard results of general relativity, serving as a consistency check of the analysis. The resulting perturbation equations reveal nontrivial modifications relative to general gravity, particularly in the scalar and tensor sectors. These additional contributions may influence the dynamics of structure formation, the evolution of gravitational waves, and the imprint of early-universe physics on the cosmic microwave background.

We started with the SVT decomposition of the perturbed tetrad through the choice \eqref{eq_52} and the perturbed energy-momentum tensor in \eqref{eq_55}. Through Eqs. \eqref{eq_61} and \eqref{eq_62}, we showed the explicit gauge dependence and presented methods of gauge-fixing and obtaining gauge-invariant variables. In Sec. \ref{sec:f_T_TG_pert}, we analyzed all the different modes of perturbations. One of the key observations from the tensor sector of the theory is that the two tensor modes propagate at the speed of light, consistent with the constraints imposed by the multimessenger events GW170817 and GRB170817A. The vector modes decay in these classes of theory, an expected behaviour in an expanding background, whereas the pseudovector modes are entirely constrained by the antisymmetric part of Eq. \eqref{eq_74}, demonstrating a well-behaved vector sector. Finally, for scalar perturbations which are of more significance, we presented equations in gauge-ready form \eqref{eq_80}--\eqref{eq_84} as well as gauge-invariant form (listed in \cite{Shivamgit2025}) in terms of gauge-invariant variables \eqref{eq_85}. Also in Sec. \ref{sec:gauge_ch}, we presented linearised scalar perturbation equations in various gauge choices that are useful in different scenarios. Additionally, pseudoscalar modes do not enter the field equations and act as remnent symmetry.

The principal motivation underlying this work was the determination of gauge-invariant perturbations of the $F(T, T_G)$ cosmology and obtaining the stability conditions avoiding Laplacian or ghost instabilities. An interesting and important direction is examining the numerical strength of the propagating degrees of freedom and their potential implications for modifications to the standard cosmological paradigm. Addressing this requires the development of further physically motivated models, which can be systematically confronted with a broad range of observational data, particularly those associated with the power spectra of perturbative modes. \skm{Another extension of the present analysis would be to investigate non-isotropic cosmological backgrounds. The current framework assumes the isotropy of the FLRW metric, however, adopting an anisotropic background (e.g., Bianchi Type I) would substantially alter the gauge-invariant perturbation formalism. In this context, such perturbations can be interpreted as effectively encoding nonlinear deviations from the FLRW background. Moreover, the background torsion scalar $T$ and the Gauss-Bonnet invariant $T_G$ would develop direction-dependent contributions, which would generically induce couplings between scalar, vector, and tensor sectors that remain decoupled in the isotropic limit.
}
\appendix
\section{Perturbative expansion}\label{app_A}
This section presents the teleparallel quantities at the background level and their first-order perturbative expansions.

\subsection{Background}

The non-zero components of the torsion tensor, torsion vector, torsion scalar, and Gauss-Bonnet term in the background (flat FLRW) are
\begin{eqnarray}
    T^{\hat{i}}\,_{0\hat{j}} & = & H\delta^{\hat{i}}{}_{\hat{j}}\,,\\
 T^\alpha\,_{0 \alpha} & = & 3H \,,\\
 T^\alpha\,_{\hat{i}\alpha}& = &0\,,\\
    T & = & 6H^{2}\,,\\
   T_G &=&  24 H^{2} \left(\dot{H} + H^{2}\right)\,.
\end{eqnarray}
In the following, we express first-order perturbations in components of the torsion tensor, torsion vector, torsion scalar, and Gauss-Bonnet scalar for different modes separately.
\subsection{Tensor perturbations}\label{ten_app}

For tensor perturbations, non-zero components of the torsion tensor, torsion vector, torsion and Gauss-Bonnet scalars are 
\begin{eqnarray}
\delta T^{\hat{i}}\,_{0\hat{j}}  & = &  \frac{1}{2}\dot h^{\hat{i}}\,_{\hat{j}}\,, \\
\delta T^{\hat{i}}\,_{\hat{j}\hat{k}}  & = &\frac{1}{2}\partial_{\hat{j}} h^{\hat{i}}\,_{\hat{k}}- \frac{1}{2}\partial_{\hat{k}} h^{\hat{i}}\,_{\hat{j}}\,,\\
\delta T^\alpha\,_{0 \alpha}  & = & 0\, ,\\
\delta T^\alpha\,_{\hat{i}\alpha} & = &0\,,\\ 
\delta T& = &0\,,\\
    \delta T_G& = &0\,.
\end{eqnarray}

\subsection{Vector and pseudovector perturbations}\label{vec_app}
The non-zero components of the vector and pseudovector perturbations for the torsion tensor, torsion vector, torsion and Gauss-Bonnet scalars are 
\begin{eqnarray}
\delta T^{0}\,_{0\hat{i}} & = & a\dot{\beta}_{\hat{i}}\,,\\
\delta T^{\hat{i}}\,_{0\hat{j}} & = & 2\partial^{\hat{i}}\dot{h}_{\hat{j}}-\frac{1}{a}\partial_{\hat{j}}B^{\hat{i}}+\epsilon^{\hat{i}}\,_{\hat{j}\hat{k}}\,\dot{\sigma}^{\hat{k}}\,,\\
\delta T^{0}\,_{\hat{i}\hat{j}} & = & a(\partial_{\hat{i}}\beta_{\hat{j}}-\partial_{\hat{j}}\beta_{\hat{i}})\,,\\
\delta T^{\hat{i}}\,_{\hat{j}\hat{k}} & = & 2(\partial^{\hat{i}}\partial_{\hat{j}}h_{\hat{k}}-\partial^{\hat{i}}\partial_{\hat{k}}h_{\hat{j}})+(\epsilon^{\hat{i}}\,_{\hat{k}\hat{l}}\,\partial_{\hat{j}}\sigma^{\hat{l}}-\epsilon^{\hat{i}}\,_{\hat{j}\hat{l}}\,\partial_{\hat{k}}\sigma^{\hat{l}})\,,\\
\delta T^\alpha\,_{0 \alpha}  & = & 0\, ,\\
\delta T^\alpha\,_{\hat{i}\alpha} & = &\epsilon_{\hat{i}\hat{j}\hat{l}}\,\partial^{\hat{j}}\sigma^{\hat{l}}- a\dot{\beta}_{\hat{i}}-2\triangle h_{\hat{i}}\,,\\
 \delta T& = &0\,,\\
    \delta T_G& = &0\,.
\end{eqnarray}
\subsection{Scalar and pseudoscalar perturbations}
\label{sca_app}
The non-zero components of the torsion tensor, torsion vector, torsion scalar and Gauss-Bonnet term for scalar and pseudoscalar perturbations are
\begin{eqnarray}
\delta T^{0}\,_{0\hat{i}} & = & \partial_{\hat{i}}(a\dot{\beta}- \phi)\,,\\
\delta T^{\hat{i}}\,_{0\hat{j}} & = & \partial^{\hat{i}}\partial_{\hat{j}}(\dot{E}-a^{-1}B)+\epsilon^{\hat{i}}\,_{\hat{j}\hat{l}}\,\partial^{\hat{l}}\dot{\sigma}+\dot{\psi}\delta^{\hat{i}}\,_{\hat{j}}\,,\\
\delta T^{0}\,_{\hat{i}\hat{j}} & = & 0\,,\\
\delta T^{\hat{i}}\,_{\hat{j}\hat{k}} & = &\delta^{\hat{i}}\,_{\hat{k}}\partial_{\hat{j}}\psi -\delta^{\hat{i}}\,_{\hat{j}}\partial_{\hat{k}}\psi+\big(\epsilon^{\hat{i}}\,_{\hat{k}\hat{m}}\,\partial_{\hat{j}}\partial^{\hat{m}}\sigma-\epsilon^{\hat{i}}\,_{\hat{j}\hat{m}}\,\partial_{\hat{k}}\partial^{\hat{m}}\sigma\big)\,,\\
\delta T^\alpha\,_{0 \alpha}  & = & \triangle (\dot{E}-a^{-1}B)+3\dot{\psi}\, ,\\
\delta T^\alpha\,_{\hat{i}\alpha} & = &\partial_{\hat{i}}(\phi-a\dot{\beta}+2\psi )\,,\\
 \delta T& = &4 H \left( \frac{-\triangle B}{a} - 3 H \phi +\triangle \dot{E} + 3 \dot{\psi} \right)\,,\\
    \delta T_G& = &  8 H \frac{\triangle}{a^{2}}\left[
    H \left( a^{2} \ddot{E} + 4 a^{2} \dot{E} H + a \dot{\beta} - 3 a B H + a \beta H - \phi \right)
    + 2 a \dot{H} (a \dot{E} - B)
    - a \dot{B} H
\right]\nonumber \\&&- 24 H \Big[
    H \left( -\ddot{\psi} + 4 H^{2} \phi + H \dot{\phi} - 4 H \dot{\psi} \right) + \dot{H} \left( 4 H \phi - 2 \dot{\psi} \right)
\Big]\,.
\end{eqnarray}

In terms of gauge invariant combinations \eqref{eq_79}, the first-order variation of scalars are

\begin{eqnarray}
 \delta T& = &12 H \left( a \beta \dot{H} - H (\mathcal{X}_1 + \mathcal{X}_2) + \dot{\mathcal{X}}_2 \right) + \frac{4 H \triangle \mathcal{X}_3}{a}\,,\\
 \delta T_G& = & 24 H \Biggl[
    a \beta H \ddot{H}
    + H \ddot{\mathcal{X}}_2
    + 2 a \beta \dot{H} (2 H^{2} + \dot{H})
    - H^{2} \left( 4 H (\mathcal{X}_1 + \mathcal{X}_2) + \dot{\mathcal{X}}_1 - 3 \dot{\mathcal{X}}_2 \right) \nonumber \\
&&
    - 4 \dot{H} H (\mathcal{X}_1 + \mathcal{X}_2)
    + 2 \dot{H} \dot{\mathcal{X}}_2
\Biggr]
- 8 H\frac{ \triangle}{a^{2}} \left[
    H \left( -3 a H \mathcal{X}_3 - a \dot{\mathcal{X}}_3 + \mathcal{X}_1 + \mathcal{X}_2 \right)
    - 2 a \dot{H} \mathcal{X}_3
\right].
\end{eqnarray}

All components of contortion and superpotential tensors can be obtained by using the torsion tensor components provided above for different modes, as the former are just linear combinations of the latter. 

\section{Matter conservation equations}\label{app_B}
\subsection{Background}
The matter components are independently conserved, yielding the standard conservation equation for a perfect fluid 
\begin{equation}
    \bar{\nabla}_{\nu}\Theta_{0}\,^{\nu}:\qquad\dot{\rho}+3H(\rho+p)=0\,.
\end{equation}
\subsection{Scalar Perturbations}
Below are first-order expansions of components of the continuity equation for scalar perturbations: 
\begin{eqnarray}
     \delta \,\bar{\nabla}_{\nu}\Theta_{0}\,^{\nu}&:&\quad\dot{\delta \rho} + 3 H (\delta p + \delta \rho) +\frac{ (p + \rho) \triangle\left( a \dot{E} + U \right)}{a} + 3 \dot{\psi} (p + \rho)=0\,,\\
      \delta \,\bar{\nabla}_{\nu}\Theta_{\hat{i}}\,^{\nu}&:&\quad \partial_{\hat{i}}\left[\delta p+ a \left(\dot{p} + \dot{\rho}\right) (B-\beta  + U)+(p + \rho) \left(4 a H (B-\beta  + U) + a \left(\dot{B}-\dot{\beta}  + \dot{U}\right) + \phi \right)\right]=0\,.
\end{eqnarray}
Throughout the above expressions, the Einstein summation convention has been employed.

\section*{Acknowledgments}
B.M. acknowledges the support of Anusandhan National Research Foundation(ANRF) for the grant (File No: CRG/2023/000475). This article is also based upon work from COST Action CA21136 Addressing observational tensions in cosmology with systematics and fundamental physics (CosmoVerse) supported by COST (European Cooperation in Science and Technology. J.L.S. would also like to acknowledge funding from ``Xjenza Malta'' as part of the ``Technology Development Programme'' DTP-2024-014 (CosmicLearning) Project, and the Project BridgingCosmology which is financed by Xjenza Malta and the Scientific Technology Research Council (TUBITAK), through the Xjenza Malta - TUBITAK 2024 Joint Call for R\&I projects. This initiative is part of the PRIMA Programme supported by the European Union. S.K.M. would like to thank Sebastian Bahamonde, The University of Tokyo, Tokyo, Japan  for fruitful discussions on field equations.

\bibliographystyle{utphys}
\bibliography{references}

\providecommand{\href}[2]{#2}\begingroup\raggedright\begin{thebibliography}{10}

\bibitem{Peebles:2002gy}
P.~J.~E. Peebles and B.~Ratra, ``{The Cosmological Constant and Dark Energy},'' \href{https://doi.org/10.1103/RevModPhys.75.559}{{\em Rev. Mod. Phys.} {\bf 75} (2003)  559--606}, \href{http://arxiv.org/abs/astro-ph/0207347}{{\tt arXiv:astro-ph/0207347}}.

\bibitem{Copeland:2006wr}
E.~J. Copeland, M.~Sami, and S.~Tsujikawa, ``{Dynamics of dark energy},'' \href{https://doi.org/10.1142/S021827180600942X}{{\em Int. J. Mod. Phys. D} {\bf 15} (2006)  1753--1936}, \href{http://arxiv.org/abs/hep-th/0603057}{{\tt arXiv:hep-th/0603057}}.

\bibitem{Riess:1998cb}
{\bf Supernova Search Team} Collaboration, A.~G. Riess {\em et al.}, ``{Observational evidence from supernovae for an accelerating universe and a cosmological constant},'' \href{https://doi.org/10.1086/300499}{{\em Astron. J.} {\bf 116} (1998)  1009--1038}, \href{http://arxiv.org/abs/astro-ph/9805201}{{\tt arXiv:astro-ph/9805201}}.

\bibitem{Perlmutter:1998np}
{\bf Supernova Cosmology Project} Collaboration, S.~Perlmutter {\em et al.}, ``{Measurements of $\Omega$ and $\Lambda$ from 42 High Redshift Supernovae},'' \href{https://doi.org/10.1086/307221}{{\em Astrophys. J.} {\bf 517} (1999)  565--586}, \href{http://arxiv.org/abs/astro-ph/9812133}{{\tt arXiv:astro-ph/9812133}}.

\bibitem{Baudis:2016qwx}
L.~Baudis, ``{Dark matter detection},'' \href{https://doi.org/10.1088/0954-3899/43/4/044001}{{\em J. Phys. G} {\bf 43} (2016) no.~4, 044001}.

\bibitem{Bertone:2004pz}
G.~Bertone, D.~Hooper, and J.~Silk, ``{Particle dark matter: Evidence, candidates and constraints},'' \href{https://doi.org/10.1016/j.physrep.2004.08.031}{{\em Phys. Rept.} {\bf 405} (2005)  279--390}, \href{http://arxiv.org/abs/hep-ph/0404175}{{\tt arXiv:hep-ph/0404175}}.

\bibitem{Weinberg:1988cp}
S.~Weinberg, ``{The Cosmological Constant Problem},'' \href{https://doi.org/10.1103/RevModPhys.61.1}{{\em Rev. Mod. Phys.} {\bf 61} (1989)  1--23}.

\bibitem{CosmoVerseNetwork:2025alb}
{\bf CosmoVerse Network} Collaboration, E.~Di~Valentino {\em et al.}, ``{The CosmoVerse White Paper: Addressing observational tensions in cosmology with systematics and fundamental physics},'' \href{https://doi.org/10.1016/j.dark.2025.101965}{{\em Phys. Dark Univ.} {\bf 49} (2025)  101965}, \href{http://arxiv.org/abs/2504.01669}{{\tt arXiv:2504.01669 [astro-ph.CO]}}.

\bibitem{DiValentino:2020vhf}
E.~Di~Valentino {\em et al.}, ``{Snowmass2021 - Letter of interest cosmology intertwined I: Perspectives for the next decade},'' \href{https://doi.org/10.1016/j.astropartphys.2021.102606}{{\em Astropart. Phys.} {\bf 131} (2021)  102606}, \href{http://arxiv.org/abs/2008.11283}{{\tt arXiv:2008.11283 [astro-ph.CO]}}.

\bibitem{DiValentino:2020zio}
E.~Di~Valentino {\em et al.}, ``{Snowmass2021 - Letter of interest cosmology intertwined II: The hubble constant tension},'' \href{https://doi.org/10.1016/j.astropartphys.2021.102605}{{\em Astropart. Phys.} {\bf 131} (2021)  102605}, \href{http://arxiv.org/abs/2008.11284}{{\tt arXiv:2008.11284 [astro-ph.CO]}}.

\bibitem{DiValentino:2020vvd}
E.~Di~Valentino {\em et al.}, ``{Cosmology Intertwined III: $f \sigma_8$ and $S_8$},'' \href{https://doi.org/10.1016/j.astropartphys.2021.102604}{{\em Astropart. Phys.} {\bf 131} (2021)  102604}, \href{http://arxiv.org/abs/2008.11285}{{\tt arXiv:2008.11285 [astro-ph.CO]}}.

\bibitem{Staicova:2021ajb}
D.~Staicova, \href{https://doi.org/10.1142/9789811269776_0151}{``{Hints for the H0 {\textemdash} rd tension in uncorrelated Baryon Acoustic Oscillations dataset},''} in {\em {16th Marcel Grossmann Meeting on~Recent Developments in Theoretical and Experimental General Relativity, Astrophysics and Relativistic Field Theories}}.
\newblock 11, 2021.
\newblock \href{http://arxiv.org/abs/2111.07907}{{\tt arXiv:2111.07907 [astro-ph.CO]}}.

\bibitem{LUX:2016ggv}
{\bf LUX} Collaboration, D.~S. Akerib {\em et al.}, ``{Results from a search for dark matter in the complete LUX exposure},'' \href{https://doi.org/10.1103/PhysRevLett.118.021303}{{\em Phys. Rev. Lett.} {\bf 118} (2017) no.~2, 021303}, \href{http://arxiv.org/abs/1608.07648}{{\tt arXiv:1608.07648 [astro-ph.CO]}}.

\bibitem{Gaitskell:2004gd}
R.~J. Gaitskell, ``{Direct detection of dark matter},'' \href{https://doi.org/10.1146/annurev.nucl.54.070103.181244}{{\em Ann. Rev. Nucl. Part. Sci.} {\bf 54} (2004)  315--359}.

\bibitem{Feng:2010gw}
J.~L. Feng, ``{Dark Matter Candidates from Particle Physics and Methods of Detection},'' \href{https://doi.org/10.1146/annurev-astro-082708-101659}{{\em Ann. Rev. Astron. Astrophys.} {\bf 48} (2010)  495--545}, \href{http://arxiv.org/abs/1003.0904}{{\tt arXiv:1003.0904 [astro-ph.CO]}}.

\bibitem{Dodelson:1993je}
S.~Dodelson and L.~M. Widrow, ``{Sterile-neutrinos as dark matter},'' \href{https://doi.org/10.1103/PhysRevLett.72.17}{{\em Phys. Rev. Lett.} {\bf 72} (1994)  17--20}, \href{http://arxiv.org/abs/hep-ph/9303287}{{\tt arXiv:hep-ph/9303287}}.

\bibitem{Joyce:2014kja}
A.~Joyce, B.~Jain, J.~Khoury, and M.~Trodden, ``{Beyond the Cosmological Standard Model},'' \href{https://doi.org/10.1016/j.physrep.2014.12.002}{{\em Phys. Rept.} {\bf 568} (2015)  1--98}, \href{http://arxiv.org/abs/1407.0059}{{\tt arXiv:1407.0059 [astro-ph.CO]}}.

\bibitem{Abazajian:2012ys}
K.~N. Abazajian {\em et al.}, ``{Light Sterile Neutrinos: A White Paper},'' \href{http://arxiv.org/abs/1204.5379}{{\tt arXiv:1204.5379 [hep-ph]}}.

\bibitem{Benisty:2021gde}
D.~Staicova and D.~Benisty, ``{Constraining the dark energy models using baryon acoustic oscillations: An approach independent of H0 {\ensuremath{\cdot}} rd},'' \href{https://doi.org/10.1051/0004-6361/202244366}{{\em Astron. Astrophys.} {\bf 668} (2022)  A135}, \href{http://arxiv.org/abs/2107.14129}{{\tt arXiv:2107.14129 [astro-ph.CO]}}.

\bibitem{Benisty:2020otr}
D.~Benisty and D.~Staicova, ``{Testing late-time cosmic acceleration with uncorrelated baryon acoustic oscillation dataset},'' \href{https://doi.org/10.1051/0004-6361/202039502}{{\em Astron. Astrophys.} {\bf 647} (2021)  A38}, \href{http://arxiv.org/abs/2009.10701}{{\tt arXiv:2009.10701 [astro-ph.CO]}}.

\bibitem{Bamba:2012cp}
K.~Bamba, S.~Capozziello, S.~Nojiri, and S.~D. Odintsov, ``{Dark energy cosmology: the equivalent description via different theoretical models and cosmography tests},'' \href{https://doi.org/10.1007/s10509-012-1181-8}{{\em Astrophys. Space Sci.} {\bf 342} (2012)  155--228}, \href{http://arxiv.org/abs/1205.3421}{{\tt arXiv:1205.3421 [gr-qc]}}.

\bibitem{Clifton:2011jh}
T.~Clifton, P.~G. Ferreira, A.~Padilla, and C.~Skordis, ``{Modified Gravity and Cosmology},'' \href{https://doi.org/10.1016/j.physrep.2012.01.001}{{\em Phys. Rept.} {\bf 513} (2012)  1--189}, \href{http://arxiv.org/abs/1106.2476}{{\tt arXiv:1106.2476 [astro-ph.CO]}}.

\bibitem{CANTATA:2021ktz}
{\bf CANTATA} Collaboration, E.~N. Saridakis, R.~Lazkoz, V.~Salzano, P.~Vargas~Moniz, S.~Capozziello, J.~Beltr{\'a}n~Jim{\'e}nez, M.~De~Laurentis, and G.~J. Olmo, eds., \href{https://doi.org/10.1007/978-3-030-83715-0}{{\em {Modified Gravity and Cosmology. An Update by the CANTATA Network}}}.
\newblock Springer, 2021.
\newblock \href{http://arxiv.org/abs/2105.12582}{{\tt arXiv:2105.12582 [gr-qc]}}.

\bibitem{Bahamonde_2023}
S.~Bahamonde, K.~F. Dialektopoulos, C.~Escamilla-Rivera, {\em et al.}, ``Teleparallel gravity: from theory to cosmology,'' \href{https://doi.org/10.1088/1361-6633/ac9cef}{{\em Reports on Progress in Physics} {\bf 86} (2023) no.~2, 026901}. \url{http://dx.doi.org/10.1088/1361-6633/ac9cef}.

\bibitem{AlvesBatista:2021gzc}
R.~Alves~Batista {\em et al.}, ``{EuCAPT White Paper: Opportunities and Challenges for Theoretical Astroparticle Physics in the Next Decade},'' \href{http://arxiv.org/abs/2110.10074}{{\tt arXiv:2110.10074 [astro-ph.HE]}}.

\bibitem{Addazi:2021xuf}
A.~Addazi {\em et al.}, ``{Quantum gravity phenomenology at the dawn of the multi-messenger era{\textemdash}A review},'' \href{https://doi.org/10.1016/j.ppnp.2022.103948}{{\em Prog. Part. Nucl. Phys.} {\bf 125} (2022)  103948}, \href{http://arxiv.org/abs/2111.05659}{{\tt arXiv:2111.05659 [hep-ph]}}.

\bibitem{Capozziello:2011et}
S.~Capozziello and M.~De~Laurentis, ``{Extended Theories of Gravity},'' \href{https://doi.org/10.1016/j.physrep.2011.09.003}{{\em Phys. Rept.} {\bf 509} (2011)  167--321}, \href{http://arxiv.org/abs/1108.6266}{{\tt arXiv:1108.6266 [gr-qc]}}.

\bibitem{Krssak:2018ywd}
M.~Krssak, R.~J. van~den Hoogen, J.~G. Pereira, C.~G. B{\"o}hmer, and A.~A. Coley, ``{Teleparallel theories of gravity: illuminating a fully invariant approach},'' \href{https://doi.org/10.1088/1361-6382/ab2e1f}{{\em Class. Quant. Grav.} {\bf 36} (2019) no.~18, 183001}, \href{http://arxiv.org/abs/1810.12932}{{\tt arXiv:1810.12932 [gr-qc]}}.

\bibitem{Cai:2015emx}
Y.-F. Cai, S.~Capozziello, M.~De~Laurentis, and E.~N. Saridakis, ``{f(T) teleparallel gravity and cosmology},'' \href{https://doi.org/10.1088/0034-4885/79/10/106901}{{\em Rept. Prog. Phys.} {\bf 79} (2016) no.~10, 106901}, \href{http://arxiv.org/abs/1511.07586}{{\tt arXiv:1511.07586 [gr-qc]}}.

\bibitem{Aldrovandi:2013wha}
R.~Aldrovandi and J.~G. Pereira, \href{https://doi.org/10.1007/978-94-007-5143-9}{{\em {Teleparallel Gravity}}}, vol.~173.
\newblock Springer, Dordrecht,
2013.
\newblock
%%CITATION = FTPHD,173,;%%.

\bibitem{Maluf:2013gaa}
J.~W. Maluf, ``{The teleparallel equivalent of general relativity},'' \href{https://doi.org/10.1002/andp.201200272}{{\em Annalen Phys.} {\bf 525} (2013)  339--357}, \href{http://arxiv.org/abs/1303.3897}{{\tt arXiv:1303.3897 [gr-qc]}}.

\bibitem{aldrovandi1995introduction}
R.~Aldrovandi and J.~Pereira, {\em An Introduction to Geometrical Physics}.
\newblock World Scientific, 1995.
\newblock \url{https://books.google.com.mt/books?id=w8hBT4DV1vkC}.

\bibitem{Ferraro:2006jd}
R.~Ferraro and F.~Fiorini, ``{Modified teleparallel gravity: Inflation without inflaton},'' \href{https://doi.org/10.1103/PhysRevD.75.084031}{{\em Phys. Rev. D} {\bf 75} (2007)  084031}, \href{http://arxiv.org/abs/gr-qc/0610067}{{\tt arXiv:gr-qc/0610067}}.

\bibitem{Ferraro:2008ey}
R.~Ferraro and F.~Fiorini, ``{On Born-Infeld Gravity in Weitzenbock spacetime},'' \href{https://doi.org/10.1103/PhysRevD.78.124019}{{\em Phys. Rev. D} {\bf 78} (2008)  124019}, \href{http://arxiv.org/abs/0812.1981}{{\tt arXiv:0812.1981 [gr-qc]}}.

\bibitem{Bengochea:2008gz}
G.~R. Bengochea and R.~Ferraro, ``{Dark torsion as the cosmic speed-up},'' \href{https://doi.org/10.1103/PhysRevD.79.124019}{{\em Phys. Rev. D} {\bf 79} (2009)  124019}, \href{http://arxiv.org/abs/0812.1205}{{\tt arXiv:0812.1205 [astro-ph]}}.

\bibitem{Linder:2010py}
E.~V. Linder, ``{Einstein's Other Gravity and the Acceleration of the Universe},'' \href{https://doi.org/10.1103/PhysRevD.81.127301}{{\em Phys. Rev. D} {\bf 81} (2010)  127301}, \href{http://arxiv.org/abs/1005.3039}{{\tt arXiv:1005.3039 [astro-ph.CO]}}. [Erratum: Phys.Rev.D 82, 109902 (2010)].

\bibitem{Chen:2010va}
S.-H. Chen, J.~B. Dent, S.~Dutta, and E.~N. Saridakis, ``{Cosmological perturbations in f(T) gravity},'' \href{https://doi.org/10.1103/PhysRevD.83.023508}{{\em Phys. Rev. D} {\bf 83} (2011)  023508}, \href{http://arxiv.org/abs/1008.1250}{{\tt arXiv:1008.1250 [astro-ph.CO]}}.

\bibitem{Bahamonde:2019zea}
S.~Bahamonde, K.~Flathmann, and C.~Pfeifer, ``{Photon sphere and perihelion shift in weak $f(T)$ gravity},'' \href{https://doi.org/10.1103/PhysRevD.100.084064}{{\em Phys. Rev. D} {\bf 100} (2019) no.~8, 084064}, \href{http://arxiv.org/abs/1907.10858}{{\tt arXiv:1907.10858 [gr-qc]}}.

\bibitem{Paliathanasis:2017htk}
A.~Paliathanasis, J.~Levi~Said, and J.~D. Barrow, ``{Stability of the Kasner Universe in f(T) Gravity},'' \href{https://doi.org/10.1103/PhysRevD.97.044008}{{\em Phys. Rev. D} {\bf 97} (2018) no.~4, 044008}, \href{http://arxiv.org/abs/1709.03432}{{\tt arXiv:1709.03432 [gr-qc]}}.

\bibitem{Farrugia:2020fcu}
G.~Farrugia, J.~Levi~Said, and A.~Finch, ``{Gravitoelectromagnetism, Solar System Tests, and Weak-Field Solutions in $f (T,B)$ Gravity with Observational Constraints},'' \href{https://doi.org/10.3390/universe6020034}{{\em Universe} {\bf 6} (2020) no.~2, 34}, \href{http://arxiv.org/abs/2002.08183}{{\tt arXiv:2002.08183 [gr-qc]}}.

\bibitem{Bahamonde:2021srr}
S.~Bahamonde, A.~Golovnev, M.-J. Guzm{\'a}n, J.~L. Said, and C.~Pfeifer, ``{Black holes in f(T,B) gravity: exact and perturbed solutions},'' \href{https://doi.org/10.1088/1475-7516/2022/01/037}{{\em JCAP} {\bf 01} (2022) no.~01, 037}, \href{http://arxiv.org/abs/2110.04087}{{\tt arXiv:2110.04087 [gr-qc]}}.

\bibitem{Bahamonde:2020bbc}
S.~Bahamonde, J.~Levi~Said, and M.~Zubair, ``{Solar system tests in modified teleparallel gravity},'' \href{https://doi.org/10.1088/1475-7516/2020/10/024}{{\em JCAP} {\bf 10} (2020)  024}, \href{http://arxiv.org/abs/2006.06750}{{\tt arXiv:2006.06750 [gr-qc]}}.

\bibitem{Bahamonde:2022ohm}
S.~Bahamonde, K.~F. Dialektopoulos, M.~Hohmann, J.~Levi~Said, C.~Pfeifer, and E.~N. Saridakis, ``{Perturbations in non-flat cosmology for f(T) gravity},'' \href{https://doi.org/10.1140/epjc/s10052-023-11322-3}{{\em Eur. Phys. J. C} {\bf 83} (2023) no.~3, 193}, \href{http://arxiv.org/abs/2203.00619}{{\tt arXiv:2203.00619 [gr-qc]}}.

\bibitem{Bahamonde:2020lsm}
S.~Bahamonde, V.~Gakis, S.~Kiorpelidi, {\em et al.}, ``{Cosmological perturbations in modified teleparallel gravity models: Boundary term extension},'' \href{https://doi.org/10.1140/epjc/s10052-021-08833-2}{{\em Eur. Phys. J. C} {\bf 81} (2021) no.~1, 53}, \href{http://arxiv.org/abs/2009.02168}{{\tt arXiv:2009.02168 [gr-qc]}}.

\bibitem{Kofinas:2014daa}
G.~Kofinas and E.~N. Saridakis, ``{Cosmological applications of $F(T,T_G)$ gravity},'' \href{https://doi.org/10.1103/PhysRevD.90.084045}{{\em Phys. Rev. D} {\bf 90} (2014)  084045}, \href{http://arxiv.org/abs/1408.0107}{{\tt arXiv:1408.0107 [gr-qc]}}.

\bibitem{Kofinas:2014aka}
G.~Kofinas, G.~Leon, and E.~N. Saridakis, ``{Dynamical behavior in $f(T,T_G)$ cosmology},'' \href{https://doi.org/10.1088/0264-9381/31/17/175011}{{\em Class. Quant. Grav.} {\bf 31} (2014)  175011}, \href{http://arxiv.org/abs/1404.7100}{{\tt arXiv:1404.7100 [gr-qc]}}.

\bibitem{Kofinas:2014owa}
G.~Kofinas and E.~N. Saridakis, ``{Teleparallel equivalent of Gauss-Bonnet gravity and its modifications},'' \href{https://doi.org/10.1103/PhysRevD.90.084044}{{\em Phys. Rev. D} {\bf 90} (2014)  084044}, \href{http://arxiv.org/abs/1404.2249}{{\tt arXiv:1404.2249 [gr-qc]}}.

\bibitem{Bahamonde:2016kba}
S.~Bahamonde and C.~G. B{\"{o}}hmer, ``{Modified teleparallel theories of gravity: Gauss$-$Bonnet and trace extensions},'' \href{https://doi.org/10.1140/epjc/s10052-016-4419-8}{{\em Eur. Phys. J. C} {\bf 76} (2016) no.~10, 578},
\href{http://arxiv.org/abs/1606.05557}{{\tt arXiv:1606.05557 [gr-qc]}}.
%%CITATION = ARXIV:1606.05557;%%.

\bibitem{delaCruz-Dombriz:2017lvj}
A.~de~la Cruz-Dombriz, G.~Farrugia, J.~L. Said, and D.~Saez-Gomez, ``{Cosmological reconstructed solutions in extended teleparallel gravity theories with a teleparallel Gauss\textendash{}Bonnet term},'' \href{https://doi.org/10.1088/1361-6382/aa93c8}{{\em Class. Quant. Grav.} {\bf 34} (2017) no.~23, 235011}, \href{http://arxiv.org/abs/1705.03867}{{\tt arXiv:1705.03867 [gr-qc]}}.

\bibitem{delaCruz-Dombriz:2018nvt}
A.~de~la Cruz-Dombriz, G.~Farrugia, J.~L. Said, and D.~S\'aez-Chill\'on~G\'omez, ``{Cosmological bouncing solutions in extended teleparallel gravity theories},'' \href{https://doi.org/10.1103/PhysRevD.97.104040}{{\em Phys. Rev. D} {\bf 97} (2018) no.~10, 104040}, \href{http://arxiv.org/abs/1801.10085}{{\tt arXiv:1801.10085 [gr-qc]}}.

\bibitem{Pradhan:2023mjx}
A.~Pradhan, A.~Dixit, and S.~Gupta, ``{Parametrization of Hubble parameter in $f(T,T_G)$ gravity model with tachyon field},'' \href{https://doi.org/10.1142/S0219887824500130}{{\em Int. J. Geom. Meth. Mod. Phys.} {\bf 20} (2023) no.~14, 2450013}.

\bibitem{Balhara:2023mgj}
H.~Balhara, J.~K. Singh, Shaily, and E.~N. Saridakis, ``{Observational Constraints and Cosmographic Analysis of f(T,T$_{G}$) Gravity and Cosmology},'' \href{https://doi.org/10.3390/sym16101299}{{\em Symmetry} {\bf 16} (2024) no.~10, 1299}, \href{http://arxiv.org/abs/2312.17277}{{\tt arXiv:2312.17277 [gr-qc]}}.

\bibitem{Gupta:2024qyn}
S.~Gupta, A.~Dixit, A.~Pradhan, and K.~Ghaderi, ``{f(T, T$_{G}$) gravity theory: observational constraints for Barrow holographic dark energy with Hubble and Granda-Oliveros cutoff},'' \href{https://doi.org/10.1088/1402-4896/ad9e51}{{\em Phys. Scripta} {\bf 100} (2025) no.~1, 015035}.

\bibitem{Pradhan:2025rir}
A.~Pradhan, A.~Dixit, S.~Gupta, and S.~Krishnannair, ``{Reconstruction of the $\boldsymbol{F(T,T_{G})}$ Tsallis Holographic Dark Energy Model Based on Observational Constraints},'' \href{https://doi.org/10.1134/S0202289325700112}{{\em Grav. Cosmol.} {\bf 31} (2025) no.~2, 221--236}.

\bibitem{Dixit:2024qno}
A.~Dixit, S.~Gupta, and A.~Pradhan, ``{Reconstruction of F(T,TG) gravity model with scalar fields},'' \href{https://doi.org/10.1142/S0219887824501834}{{\em Int. J. Geom. Meth. Mod. Phys.} {\bf 21} (2024) no.~11, 2450183}.

\bibitem{Sultan:2025tws}
A.~M. Sultan, M.~Ali, S.~Rani, N.~Azhar, N.~Myrzakulov, and S.~Shaymatov, ``{Constraining Big Bang nucleosynthesis in f(T,B,TG,BG) gravity},'' \href{https://doi.org/10.1016/j.nuclphysb.2025.117023}{{\em Nucl. Phys. B} {\bf 1018} (2025)  117023}.

\bibitem{Majeed:2025tjp}
A.~Majeed, S.~Rani, N.~Azhar, M.~M. Alam, S.~Shaymatov, and A.~M. Sultan, ``{f(T,B,TG,BG) gravity tested through gravitational baryogenesis mechanisms},'' \href{https://doi.org/10.1016/j.dark.2025.101957}{{\em Phys. Dark Univ.} {\bf 48} (2025)  101957}.

\bibitem{TheLIGOScientific:2017qsa}
{\bf LIGO Scientific, Virgo} Collaboration, B.~P. Abbott {\em et al.}, ``{GW170817: Observation of Gravitational Waves from a Binary Neutron Star Inspiral},'' \href{https://doi.org/10.1103/PhysRevLett.119.161101}{{\em Phys. Rev. Lett.} {\bf 119} (2017) no.~16, 161101}, \href{http://arxiv.org/abs/1710.05832}{{\tt arXiv:1710.05832 [gr-qc]}}.

\bibitem{Goldstein:2017mmi}
A.~Goldstein {\em et al.}, ``{An Ordinary Short Gamma-Ray Burst with Extraordinary Implications: Fermi-GBM Detection of GRB 170817A},'' \href{https://doi.org/10.3847/2041-8213/aa8f41}{{\em Astrophys. J. Lett.} {\bf 848} (2017) no.~2, L14}, \href{http://arxiv.org/abs/1710.05446}{{\tt arXiv:1710.05446 [astro-ph.HE]}}.

\bibitem{Kamionkowski:2015yta}
M.~Kamionkowski and E.~D. Kovetz, ``{The Quest for B Modes from Inflationary Gravitational Waves},'' \href{https://doi.org/10.1146/annurev-astro-081915-023433}{{\em Ann. Rev. Astron. Astrophys.} {\bf 54} (2016)  227--269}, \href{http://arxiv.org/abs/1510.06042}{{\tt arXiv:1510.06042 [astro-ph.CO]}}.

\bibitem{Izumi:2012qj}
K.~Izumi and Y.~C. Ong, ``{Cosmological Perturbation in $f(T)$ Gravity Revisited},'' \href{https://doi.org/10.1088/1475-7516/2013/06/029}{{\em JCAP} {\bf 06} (2013)  029}, \href{http://arxiv.org/abs/1212.5774}{{\tt arXiv:1212.5774 [gr-qc]}}.

\bibitem{Tseytlin:1982firstorder}
A.~A. Tseytlin, ``{On the first-order formalism in quantum gravity},'' \href{https://doi.org/10.1088/0305-4470/15/3/005}{{\em J. Phys. A} {\bf 15} (1982) no.~3, L105}.

\bibitem{Harst:2012tetrad}
U.~Harst and M.~Reuter, ``{The `tetrad only' theory space: nonperturbative renormalization flow and asymptotic safety},'' \href{https://doi.org/10.1007/JHEP05(2012)005}{{\em JHEP} {\bf 2012} (2012) no.~05, 005}.

\bibitem{Sadovski:2025holonomic}
G.~Sadovski, ``{About the (in)equivalence between holonomic versus non-holonomic theories of gravity},'' \href{https://doi.org/10.1142/S0219887825500045}{{\em Int. J. Geom. Meth. Mod. Phys.} {\bf 22} (2025) no.~06, 2550004}.

\bibitem{Golovnev:2017}
A.~Golovnev, T.~Koivisto, and M.~Sandstad, ``{On the covariance of teleparallel gravity theories},'' \href{https://doi.org/10.1088/1361-6382/aa7830}{{\em Class. Quant. Grav.} {\bf 34} (2017) no.~14, 145013}.

\bibitem{NGR}
K.~Hayashi and T.~Shirafuji, ``New general relativity,'' \href{https://doi.org/10.1103/PhysRevD.19.3524}{{\em Phys. Rev. D} {\bf 19} (1979) no.~12, 3524--3553}.

\bibitem{Bruni:1996im}
M.~Bruni, S.~Matarrese, S.~Mollerach, and S.~Sonego, ``{Perturbations of space-time: Gauge transformations and gauge invariance at second order and beyond},'' \href{https://doi.org/10.1088/0264-9381/14/9/014}{{\em Class. Quant. Grav.} {\bf 14} (1997)  2585--2606}, \href{http://arxiv.org/abs/gr-qc/9609040}{{\tt arXiv:gr-qc/9609040}}.

\bibitem{Malik:2008im}
K.~A. Malik and D.~Wands, ``{Cosmological perturbations},'' \href{https://doi.org/10.1016/j.physrep.2009.03.001}{{\em Phys. Rept.} {\bf 475} (2009)  1--51}, \href{http://arxiv.org/abs/0809.4944}{{\tt arXiv:0809.4944 [astro-ph]}}.

\bibitem{bardeen1980gauge}
J.~M. Bardeen, ``{Gauge Invariant Cosmological Perturbations},'' \href{https://doi.org/10.1103/PhysRevD.22.1882}{{\em Phys. Rev. D} {\bf 22} (1980)  1882--1905}.

\bibitem{Mukhanov:1990me}
V.~F. Mukhanov, H.~A. Feldman, and R.~H. Brandenberger, ``{Theory of cosmological perturbations. Part 1. Classical perturbations. Part 2. Quantum theory of perturbations. Part 3. Extensions},''
\href{https://doi.org/10.1016/0370-1573(92)90044-Z}{{\em Phys. Rept.} {\bf 215} (1992)  203--333}.
%%CITATION = PRPLC,215,203;%%.

\bibitem{heisenberg2023gaugeinvariantcosmologicalperturbationsgeneral}
L.~Heisenberg and M.~Hohmann, ``Gauge-invariant cosmological perturbations in general teleparallel gravity,'' \href{http://arxiv.org/abs/2311.05597}{{\tt arXiv:2311.05597 [gr-qc]}}.

\bibitem{Stewart:1974}
J.~M. Stewart and M.~Walker, ``{Perturbations of space-times in general relativity},'' \href{https://doi.org/10.1098/rspa.1974.0172}{{\em Proc. R. Soc. Lond. A} {\bf 341} (1974)  49--74}.

\bibitem{Mishra:2025}
S.~K. Mishra, J.~L. Said, and B.~Mishra, ``Propagating gravitational waves in teleparallel {G}auss-{B}onnet gravity,'' \href{https://doi.org/10.1103/p6l8-k963}{{\em Phys. Rev. D} {\bf 112} (2025)  064019}, \href{http://arxiv.org/abs/2505.18192}{{\tt arXiv:2505.18192 [gr-qc]}}.

\bibitem{Cai:2018rzd}
Y.-F. Cai, C.~Li, E.~N. Saridakis, and L.~Xue, ``{$f(T)$ gravity after GW170817 and GRB170817A},'' \href{https://doi.org/10.1103/PhysRevD.97.103513}{{\em Phys. Rev. D} {\bf 97} (2018) no.~10, 103513}, \href{http://arxiv.org/abs/1801.05827}{{\tt arXiv:1801.05827 [gr-qc]}}.

\bibitem{Shivamgit2025}
S.~K. Mishra, ``{T}{G}{B} {P}erturbations.'' \url{https://github.com/ShivamKumarMishra-01/TGB_Perturbations}, 2025.
\newblock [Accessed 01-09-2025].

\bibitem{Wu_2012}
Y.-P. Wu and C.-Q. Geng, ``Matter density perturbations in modified teleparallel theories,'' \href{https://doi.org/10.1007/jhep11(2012)142}{{\em Journal of High Energy Physics} {\bf 2012} (2012) no.~11, }. \url{http://dx.doi.org/10.1007/JHEP11(2012)142}.

\bibitem{Zheng_2011}
R.~Zheng and Q.-G. Huang, ``{Growth factor in $f(T)$ gravity},'' \href{https://doi.org/10.1088/1475-7516/2011/03/002}{{\em Journal of Cosmology and Astroparticle Physics} {\bf 2011} (2011) no.~03, }. \url{http://dx.doi.org/10.1088/1475-7516/2011/03/002}.

\bibitem{Clifton:2020}
T.~Clifton, C.~S. Gallagher, S.~Goldberg, and K.~A. Malik, ``{Viable gauge choices in cosmologies with nonlinear structures},'' \href{https://doi.org/10.1103/PhysRevD.101.063530}{{\em Phys. Rev. D} {\bf 101} (2020) no.~6, 063530}.

\end{thebibliography}\endgroup

\end{document}